\documentclass[aps,longbibliography,nofootinbib,showkeys]{revtex4-1}

\usepackage{amssymb}
\usepackage{amsfonts} 
\usepackage{graphicx} 
\usepackage{amsthm} 
\usepackage{amsmath} 
\usepackage{epsfig}
\usepackage{hyperref}
\usepackage{color}
\usepackage{bbold}
\usepackage{young}
\usepackage{youngtab}
\usepackage[boxsize=1.07em]{ytableau}
\usepackage{braket}

\usepackage{pgf}
\usepackage{bm}
\usepackage{bbold}
\usepackage{bbm}
\usepackage{graphicx}
\usepackage[makeroom]{cancel}

\usepackage[utf8]{inputenc}

\def\ic{\mathrm{i}}

\def \bc {\begin{center}}
\def \ec {\end{center}}
\def \bi {\begin{itemize}}
\def \ei {\end{itemize}}

\def \ba {\begin{array}}
\def \ea {\end{array}}

\def \bea {\begin{eqnarray}}
\def \eea {\end{eqnarray}}

\def\be{\begin{equation}}
\def\ee{\end{equation}}

\newcommand{\la}{\langle}
\newcommand{\ra}{\rangle}

\def\tr {\mathrm{tr}}

 \def\rmu {\mathrm{U}}

\def\rmu{\mathrm{U}}

\def\2cat{{\scriptstyle\mathrm{2CAT}}}
\def\3cat{{\scriptstyle\mathrm{3CAT}}}

 \theoremstyle{remark}
 
\definecolor{darkgreen}{rgb}{0,0.5,0}
\definecolor{darkblue}{rgb}{0.1,0.1,0.5}
\definecolor{darkred}{rgb}{0.8,0,0}
\definecolor{verylightgray}{rgb}{0.87,0.87,0.87}

\def\ic{\mathrm{i}}

\def\zb{{\bm{z}}}

\def\wb{{\bm{w}}}

\def\alphab{\bm{\alpha}}

\begin{document}

\title{Wigner quasi-probability distribution for symmetric multi-quDit systems and their generalized heat kernel}

\author{Manuel Calixto}
\email{calixto@ugr.es}
\affiliation{Department of Applied Mathematics, University of  Granada,
Fuentenueva s/n, 18071 Granada, Spain}
\affiliation{Institute Carlos I of Theoretical and Computational Physics (iC1), University of  Granada,
Fuentenueva s/n, 18071 Granada, Spain}
\author{Julio Guerrero}
\email{jguerrer@ujaen.es}
\affiliation{Department of Mathematics, University of Ja\'en, Campus Las Lagunillas s/n, 23071 Ja\'en, Spain}
\affiliation{Institute Carlos I of Theoretical and Computational Physics (iC1), University of  Granada,
Fuentenueva s/n, 18071 Granada, Spain}

\date{\today}


\begin{abstract}

For a symmetric $N$-quDit system described by a density matrix $\rho$, we construct a one-parameter $s$ family $\mathcal{F}^{(s)}_\rho$ of quasi-probability distributions  through generalized Fano multipole operators and Stratonovich-Weyl kernels. The corresponding phase space is the complex projective $\mathbb{C}P^{D-1}=\rmu(D)/\rmu(D-1)\times \rmu(1)$, related to fully symmetric irreducible representations of the unitary group $\rmu(D)$. For the particular cases $D=2$ (qubits) and $D=3$ (qutrits), we analyze the phase-space structure of  Schr\"odinger $\rmu(D)$-spin cat (parity adapted coherent) states and we provide plots of the corresponding Wigner $\mathcal{F}^{(0)}_\rho$ function. We examine the connection between non-classical behavior and the negativity of the Wigner function. We also compute the generalized heat kernel relating two quasi-probability distributions $\mathcal{F}^{(s)}_\rho$ and $\mathcal{F}^{(s')}_\rho$, with $t=(s'-s)/2$ playing the role of ``time'', together with their twisted Moyal product in terms of a trikernel. In the thermodynamic limit $N\to\infty$, we recover the usual Gaussian smoothing for $s'>s$.  A diagramatic interpretation of the phase-space construction in terms of Young tableaux is also provided.

\end{abstract}

\maketitle

\section{Introduction}
The Wigner $\mathcal{W}_\rho$ quasi-probability distribution function \cite{PhysRev.40.749}, in conjunction with the Husimi $\mathcal{Q}_\rho$ \cite{husimi1940} and the Glauber-Sudarshan $\mathcal{P}_\rho$ \cite{GlauberPhysRevLett.10.84,PhysRev.131.2766,PhysRevLett.10.277} functions,  provides a valuable phase-space representation for quantum states $\rho$, bridging the gap between classical statistical mechanics and quantum mechanics.  Thanks to the pioneering works of: Weyl \cite{Weyl1927}, Groenewold \cite{GROENEWOLD1946405},  Moyal \cite{Moyal_1949}, Stratonovich \cite{Stratonovich1957}, Agarwald and Wolf \cite{AgarwalPhysRevD.2.2161,AgarwalPhysRevD.2.2187,AgarwalPhysRevD.2.2206}, Cahill and Glauber \cite{PhysRev.177.1857,PhysRev.177.1882}, Berezin \cite{Berezin1975},  among others, we actually know an axiomatic approach (called ``Stratonovich-Weyl correspondence'') to deal with all these quasi-probability distribution functions based on the Heisenberg-Weyl (and harmonic oscillator) symmetry.  For example, expectation values of normally, anti-normally and symmetrically ordered product combinations of annihilation and creation operators can be evaluated from the corresponding characteristic functions for the $\mathcal{P}$, $\mathcal{Q}$ and $\mathcal{W}$ distributions, respectively. These works also contain the seed to extend the original construction to other symmetry groups. For example, Stratonovich's paper \cite{Stratonovich1957} already has  the germ for a Moyal theory for spin, later developed in \cite{VARILLY1989107,Klimov_2002}, and the construction of generalized phase-space distributions for spin-$j$ SU(2) invariant  systems \cite{AgarwalPhysRevA.24.2889}, based on spin-$N/2$, or ``atomic'' (for $N$ two-level indistinguishable atoms) coherent states \cite{JMRadcliffe_1971,GilmorePhysRevA.6.2211,GilmoreJMP} generalizing standard (canonical, harmonic oscillator) coherent states, introduced by Schr\"odinger \cite{Schrodinger1926} as minimal uncertainty wavepackets, formalized by Klauder \cite{KLAUDER1960123} and popularized by Glauber \cite{PhysRev.131.2766} in the field of quantum optics. Coherent states (CSs) are said to be the ``most classical'' among quantum states; not in vain they are excellent variational ground states of critical Hamiltonian quantum systems \cite{Gilmorebook} in the thermodynamic limit (infinite number $N$ of particles, large spin, etc.). CSs are labelled by phase-space points and turn out to be essential in the Stratonovich-Weyl (SW) mathematical framework. CSs for general symmetry groups, other than Heisenberg-Weyl, have been constructed in the literature; see e.g.  Klauder and Skagerstam \cite{Klauderbook}, Perelomov \cite{Perelomov}, Gilmore et al. \cite{RevModPhys.62.867} and Gazeau \cite{Gazeaubook} traditional books and reviews. The concept of generalized CSs and the method of harmonic analysis led Brif and Man  \cite{Brif_1998,Brif_1998,PhysRevA.59.971}  propose a construction of a one parameter $s$ family $\mathcal{F}^{(s)}_\rho$ of phase-space  functions satisfying the SW correspondence for physical systems possessing Lie-group $G$ symmetries (see also \cite{Ali2000} for a connection with wavelet theory). Wigner and other  functions have been constructed for $G=\mathrm{SU}(2)$ invariant systems \cite{AgarwalPhysRevA.24.2889,VARILLY1989107,Klimov_2002}, with  extensions to $\mathrm{SU}(3)$ \cite{MartinsKlimovGuise_SU3} and  $\mathrm{SU}(N)$   \cite{Klimov_2010}, non-compact $\mathrm{SU}(1,1)$ \cite{Klimov_2021}, 
finite systems \cite{Atakishiyev1998-2} and arbitrary quantum systems \cite{TilmaPhysRevLett.117.180401}, to name a few of them. 

In this article we build a one-parameter $s$ family $\mathcal{F}^{(s)}_\rho$ of quasi-probability distributions  for symmetric $N$-quDit systems (namely, $N$ indistinguishable $D$-level atoms). The construction is done by generalizing $\rmu(2)$ Fano multipole operators and Stratonovich-Weyl kernels to $\rmu(D)$. The phase space in this case is the complex projective space $\mathbb{C}P^{D-1}= \rmu(D)/\rmu(D-1)\times \rmu(1)$ generalizing the Bloch sphere $\mathbb{S}^2=\mathbb{C}P^{1}= \rmu(2)/\rmu(1)\times\rmu(1)$  for symmetric $N$-qubit (spin $j=N/2$) systems. The $\rmu(D)$-spin CSs are then parametrized by complex vectors $\zb=(z_1,\dots,z_{D-1})\in \mathbb{C}P^{D-1}$. 

As a nontrivial example, we analyze the phase-space structure of $\rmu(D)$-spin 
CSs adapted to parity, by computing their Wigner function $\mathcal{W}_\rho=\mathcal{F}^{(0)}_\rho$. The parity adaptation of CSs are generally named in the literature as ``Schr\"odinger cat states'' after  \cite{Dodonovcat,M.Nieto&D.Traux,V.Buzek&A.Viiella-Barranco&P.Knight,M.Hillery,Mankocat}, among others, since they are  quantum superpositions of weakly-overlapping (macroscopically distinguishable) 
quasi-classical coherent wave packets. Their non-classical properties, like squeezing and high quantum correlations, make them an important resource for quantum computation, information, communication and  metrology \cite{RevModPhys.90.035005-Pezze}.  Cat states can be generated 
via amplitude dispersion by evolving CSs in  Kerr media, with a strong nonlinear interaction. This also can be done for ``atomic Schr\"odinger cat states'' \cite{Agarwal_PhysRevA.56.2249} or  spin-squeezed states \cite{Kitagawa} in symmetric multiqubit systems \cite{Calixto_2017}.

Phase space analysis (entanglement, squeezing, Husimi function, etc) of symmetric multi-quDit sysmtems has been done in \cite{nuestroPRE,QIP-2021-Entanglement,Guerrero22,Guerrero_2023,Mayorgas2023}, with applications to quantum phase transitions in $D$-level Lipkin-Meshkov-Glick models. Here we introduce new  quasi-distribution functions in phase space  (in addition to the Husimi function) for $N$-quDit-symmetric  systems, in the hope that they will constitute an important tool for analysing their intrinsic quantum properties.  

The organization of the article is as follows. In Section \ref{cohsec} we introduce notation by reminding the definition of $\rmu(D)$ CSs on complex proyective spaces $\mathbb{C}P^{D-1}$ and their closure relations. In Section \ref{laplaciansec} we analyze the problem of Laplacian eigenfunctions on $\mathbb{C}P^{D-1}$ and the reproducing kernel. These constructions are essential tools to define generalized Fano multipole operators and the Stratonovich-Weyl kernel in Section  \ref{FanoSWsec} and the mapping of a general density matrix $\rho$ onto a $s$-parameter family of quasi-probability distribution functions $\mathcal{F}^{(s)}_\rho$. We exemplify these constructions with the particular cases $D=2$ (symmetric multi-qudits) and $D=3$ (symmetric multi-qutrits) by analyzing the phase-space structure of even and odd cat states. In Section \ref{Youngsec} we provide a diagramatic picture of our construction in terms of Young tableaux. In Section  \ref{heatsec} we study the generalized heat kernel connecting two quasi-probability distribution functions $\mathcal{F}^{(s)}_\rho$ and  $\mathcal{F}^{(s')}_\rho$, interpreted as a generalized Gaussian smoothing after an interesting relation proved in Appendix  \ref{apptaulambda}. In Sec. \ref{twistsec} we study the generalized twisted (star) product, defined through a trikernel, whose differential expression identifies the Poisson bracket as the leading term of the expansion $1/N$ (for large $N$) of the Moyal bracket. Section \ref{conclusec} is left for conclusions and outlook.

\section{Coherent states of symmetric multi-quDits}\label{cohsec}

A $D$-level $\{|i\ra,\. i=0,\dots,D-1\}$ atom/particle  carries a quDit of information. Let us think of $|i\ra$ as atom energy $E_i$ levels with $E_i<E_j$ for $i<j$. A general quDit state can be written as a linear expansion $|\bm{\zeta}\ra=\sum_{i=0}^{D-1} \zeta_i|i\ra/\|\bm{\zeta}\|$ with coefficients  $\bm{\zeta}\in \mathbb{C}^D$ and $\|\bm{\zeta}\|=(|\zeta_0|^2+\dots+|\zeta_{D-1}|^2)^{1/2}$ its norm. The probability of finding the quDit in level $i$ is then $|\zeta_i|^2/\|\bm{\zeta}\|$. We shall consider the ``low excitation regime"   in which level $i=0$ is always populated, that is, $\zeta_0\not=0$. This is just a convention related to the use of a particular patch of the complex projective space $\mathbb{C}P^{D-1}= \mathbb{C}^D/\sim$, with $\bm{\zeta}\sim \eta \bm{\zeta}, \forall \eta\in \mathbb{C}$, the equivalence relation. Chosing $\eta=1/\zeta_0$, we denote the class representative by the complex \emph{column}  vector $\zb=(z_1,\dots,z_{D-1})^T=(\zeta_1/\zeta_0,\dots, \zeta_{D-1}/\zeta_0)^T\in \mathbb{C}^{D-1}$ ($T$ means transpose). In this regime/convention a 1-quDit coherent state is written as
\be
|\zb\ra=\frac{|0\ra+z_1|1\ra+\dots+z_{D-1}|D-1\ra}{\sqrt{1+|z_1|^2+\dots+|z_{D-1}|^2}}.\label{1quDit}
\ee
For $D=2$,  the 1-qubit coherent state can be written as $|z\ra=\cos({\theta}/{2})|0\ra+\sin({\theta}/{2})e^{\ic \phi}|1\ra$, with $z=\tan(\theta/2)e^{\ic \phi}$ the sthereographic projection of a point $(\theta,\phi)$ of the Bloch sphere $\mathbb{C}P^1=\mathbb{S}^2$ onto the complex plane $\mathbb{C}\ni z$.

Of special importance will be the squared norm of the coherent state overlap
\be
Q(\zb,\wb)=|\la \wb|\zb\ra|^2=\frac{(1+\wb^\dag\zb)(1+\zb^\dag\wb)}{(1+\zb^\dag\zb)(1+\wb^\dag\wb)},\label{csoverlap}
\ee
where $\zb^\dag$ means transpose conjugate. The coherent state of $N$ symmetric quDits is the Kronecker product 
\be
 |N,\zb\ra\equiv |\zb\ra^{\otimes N}=|\zb\ra\otimes\stackrel{N}{\dots}\otimes|\zb\ra.
\ee
They can also be expanded  in terms of Fock states  ($|\vec{0}\ra$ denotes the Fock vacuum and $n_i$ the occupancy number of level $i$)
\be
|\vec{n}\ra=|n_0,\dots, n_{D-1}\ra=
\frac{(a_0^\dag)^{n_0}\dots(a_{D-1}^\dag)^{n_{D-1}}}{\sqrt{n_0!\dots n_{D-1}!}}|\vec{0}\ra, \quad \|\vec{n}\|_1=n_0+\dots+n_{D-1}=N, \label{symmetricbasis}
\ee
of $D$ bosonic modes  as the Bose-Eintein condensate 
\be
 |N,\zb\ra=\frac{1}{\sqrt{N!}}\left(
	\frac{a_0^\dag+z_1a_1^\dag+\cdots+z_{D-1} a_{D-1}^\dag}{\sqrt{1+|z_1|^2+\cdots+|z_{D-1}|^2}}\right)^{N}|\vec{0}\ra \,. \label{cohD}
\ee
The $N$-particle squared norm coherent state overlap  is simply the $N$-th power of 
\be
|\la N,\wb|N,\zb\ra|^2=|\la \wb|\zb\ra|^{2N}=Q^N(\zb,\wb).\label{csoverlapN}
\ee
$D$-mode boson operators $a_i,a_i^\dag$  annihilate and create $D$-level atoms in the $i$-th level, respectively.   $\rmu(D)$-spin operators 
\be
S_{ij}=a^\dag_i a_j, \; i,j=0,\dots,D-1\,,\label{UDgen}
\ee
conserve the total number $N$ of quDits [the linear Casimir operator $C_1=\sum_{i=0}^{D-1} S_{ii}$ of $\rmu(D)$].  The Hilbert space ${\cal H}_N$ of our symmetric  $N$-quDit system is then the  $d_{N}=\binom{N+D-1}{N}$-dimensional carrier space of the fully symmetric irreducible  representation of $\rmu(D)$. The quadratic Casimir operator  of $\rmu(D)$ 
\be
 C_2=\sum_{i,j=0}^{D-1} S_{ij}S_{ji} \label{Casimir2}
 \ee
 is then fixed to $\lambda_{N}=N(N+D-1)$. 
 
 For $D=2$-level systems (qubits),  $\rmu(2)$-spin or ``atomic'' CSs  \cite{JMRadcliffe_1971,GilmorePhysRevA.6.2211} are commonly  written in terms of the $d_{N}=\binom{N+1}{N}=N+1$ Dicke (angular momentum $j=N/2$) states $$|j,m\ra\equiv |\vec{n}\ra= |n_0=j-m,n_1=j+m\ra, \;m=-j,\dots,j,$$ as
\be
|N=2j,z\ra=\frac{1}{\sqrt{(2j)!}}\left(
	\frac{a_0^\dag+z a_1^\dag}{\sqrt{1+|z|^2}}\right)^{2j}|\vec{0}\ra =\sum_{m=-j}^j \binom{2j}{j+m}^{1/2} \frac{z^{j+m}}{(1+|z|^2)^{j}}|j,m\ra \label{coh2}
\ee
In particular, for $N=1$ and $z=\tan(\theta/2)e^{\ic\phi}$, we recover the one-qubit CS  \eqref{1quDit}  as a spin $1/2$ state 
$$|1,z\ra=\cos({\theta}/{2})\left|\frac{1}{2},-\frac{1}{2}\right\ra+\sin({\theta}/{2})e^{\ic \phi}\left|\frac{1}{2},\frac{1}{2}\right\ra.$$ 
The usual angular momentum ladder operators are $J_+=S_{10}, J_-=S_{01}$, together with $J_z=(S_{11}-S_{00})/2$. The relation between $\rmu(2)$ Casimir operators $C_1, C_2$ and $\vec{J}^2$ is $\vec{J}^2=(C_2+C_1)/4$.

CSs \eqref{cohD} form an overcomplete system for the Hilbert space $\mathcal{H}_N$ with resolution of the identity 
\be
\hat{I}=\int_{\mathbb{C}^{D-1}}d\mu_N(\zb)|N,\zb\ra\la N,\zb|.\label{resolution}
\ee
The integration measure $d\mu_N(\zb)=d_Nd\mu_0(\zb)$, with $d_N=\tbinom{N+D-1}{N}$, is a multiple of  the Haar invariant measure 
\be
d\mu_0(\zb)=\frac{(D-1)!}{\pi^{D-1}}\;\frac{\prod_{i=1}^{D-1}d\Re(z_i)d\Im(z_i)}{(1+\zb^\dag\zb)^D},\label{dmu0}
\ee
which is normalized according to $\int_{\mathbb{C}^{D-1}}d\mu_0(\zb)=1$ ($\Re$ and $\Im$ denote real an imaginary parts).

The Husimi quasi-probability distribution function of a general state $|\psi\ra\in \mathcal{H}_N$ is defined as $\mathcal{Q}_\psi(\zb)=|\la \zb|\psi\ra|^2$. Note that, in particular,  the Husimi function $\mathcal{Q}_{|N,\wb\ra}(\zb)$ of the CS $|N,\wb\ra$ coincides with the squared norm CS overlap $Q^N(\zb,\wb)$ in \eqref{csoverlapN}. In this paper, we shall recover the Husimi function $\mathcal{Q}_\rho$ from a one-parameter  $s$ family of phase-space quasi-probability distribution functions $\mathcal{F}^{(s)}_\rho$ as the case $s=-1$. Before, we need to introduce some previous material.

\section{Laplacian eigenstates and reproducing kernel}\label{laplaciansec}

The Laplace-Beltrami (Laplacian) operator on the complex projective space $\mathbb{C}P^{D-1}$ is 
\be
\Delta_\zb=(1+\zb^\dag\zb)\sum_{i,j=1}^{D-1}(\delta_{ij}+\bar{z}_iz_j)\frac{\partial^2}{\partial \bar{z}_i\partial z_j}.\label{Laplacian}
\ee
It can be seen as the differential operator version of (minus) the quadratic Casimir  \eqref{Casimir2}. We look for $\lambda$-eigenfunctions $\Psi$ of the Laplacian, that is, $\Delta_\zb\Psi(\zb)=-\lambda \Psi(\zb)$. Analiticity of  solutions restrict Laplacian eigenvalues $\lambda$ to the Casimir  \eqref{Casimir2} eigenvalues $\lambda_{N}=N(N+D-1)$. Noticing that CS overlap functions \eqref{csoverlapN} fulfill 
\be
\Delta_\zb Q^N(\zb,\wb)=-\lambda_{N}Q^N(\zb,\wb)+N^2 Q^{N-1}(\zb,\wb),
\ee
we can try Laplacian solutions of the form 
\be
K_N(\zb,\wb)=\sum_{n=0}^\infty c_{N,n}\,Q^n(\zb,\wb),\label{lambdakernel}
\ee
which, for  $\lambda_{N}=N(N+D-1)$, gives (see Ref. \cite{Qikeng})
\be
 c_{N,n}=-\frac{ \lambda_{N}^2 \, (1-N)_{n-1} (D+N)_{n-1}}{(n!)^2}c_0, \, n=1,2,\dots, N,\quad c_{N,0}=c_0\lambda_{N},
 \ee
 where $(q)_n=q(q+1)\dots (q+n-1)$ are Pochhammer symbols. Note that $ c_{N,n}=0$ for $n>N$ so that the series \eqref{lambdakernel} truncates and gives 
 \be
 K_N(\zb,\wb)= c_0 \,\lambda_{N} \, _2F_1(-N,D+N-1;1;Q(\zb,\wb)),\label{lambdakernel2}
 \ee
 in terms of hypergeometric functions $_2F_1(\alpha,\beta;\gamma;x)$, which can also be written as Jacobi polynomials $P_N^{(0,D-2)}(1-2 x)=\, _2F_1(-N,D+N-1;1;x)$, reducing to  Legendre polynomials $P_N(1-2 x)=\, _2F_1(-N,N+1;1;x)$ for the particular case of qubits ($D=2$). 
 
 The Hilbert space $\tilde{\mathcal{H}}_N$  of $\lambda_{N}$-eigenfunctions of $\Delta_\zb$ is spanned by  (see Ref. \cite{BergerBook})
 \be
 \tilde{d}_{N}=\sum_{n=0}^N d_{n}=d_{N}^2-d_{N-1}^2\label{dimtilde}
 \ee
 orthonormal basis harmonic functions $\{\Psi^{(N)}_j, j=1,\dots,\tilde{d}_{N}\} $  with respect to the Haar measure \eqref{dmu0}. The $\lambda_{N}$-kernel $K_N(\zb,\wb)$ can be expanded as
 \be
 K_N(\zb,\wb)=\sum_{j=1}^{\tilde{d}_{N}} \Psi^{(N)}_j(\bar{\wb})\Psi^{(N)}_j({\zb})=\sum_{j=1}^{\tilde{d}_{N}} \Psi^{(N)}_j(\bar{\zb})\Psi^{(N)}_j({\wb}),\label{kernelexpansion}
 \ee
 and it is a reproducing kernel 
 \be
 \Psi(\zb)=\int_{\mathbb{C}^{D-1}}K_N(\zb,\wb)\Psi(\wb)d\mu_0(\wb),
 \ee
 Since the basis $\{\Psi^{(N)}_j\}$ is orthonormal, the arbitrary constant $c_0$ of $K_N(\zb,\wb)$ in \eqref{lambdakernel2} 
is fixed according to
\be
\int_{\mathbb{C}^{D-1}}K_N(\zb,\zb)d\mu_0(\zb)=\tilde{d}_{N}\Rightarrow c_0=\tilde{d}_{N}/[ _2F_1(-N,D+N-1;1;1)\lambda_{N} ],
\ee
and therefore, the normalized $\lambda_{N}$-kernel $K_N(\zb,\wb)$ finally writes as
\be
 K_N(\zb,\wb)= \tilde{d}_{N}\,\frac{ _2F_1(-N,D+N-1;1;|\la \zb|\wb\ra|^2)}{ _2F_1(-N,D+N-1;1;1)}.\label{lambdakerneln}
 \ee
The relation \eqref{lambdakernel} can be inverted as
\be
Q^N(\zb,\wb)=\sum_{n=0}^N \tilde{c}_{N,n}\,K_n(\zb,\wb).\label{CGexpansion}
\ee
The coefficients $\tilde{c}_{N,n}$ can be calculated by making use of the orthogonality properties of Jacobbi polynomials
\be
\int_{-1}^{1} (1+y)^{D-2}P_n^{(0,D-2)}(y)P_m^{(0,D-2)}(y) \, dy=\frac{2^{D-1}}{2n+D-1}\delta_{n,m}, 
\ee
and that 
\be
\int_{-1}^{1} (y+1)^{D-2} \left(\frac{1-y}{2}\right)^N P_n^{(0,D-2)}(y)\,dy=\frac{(-1)^n 2^{D-1} N! (n+1)_{D-2}}{(N-n)! (N+1)_{D+n-1}},
\ee
which then gives
\be
\tilde{c}_{N,n}=\frac{(-1)^n N! (D+2 n-1)  (n+1)_{D-2}P_n^{(0,D-2)}(-1)}{(N-n)!  (N+1)_{D+n-1} \tilde{d}_{N}}=\frac{(D-1)!(N!)^2}{(N-n)!(N+n+D-1)!}.\label{tildec}
\ee
Writing $Q^N(\zb,\wb)=\la \wb|\zb\ra^{N} \overline{\la \wb|\zb\ra^{N} }$, we can guess from \eqref{CGexpansion} that the  coefficients $\tilde{c}_{N,n}$ are related to the Clebsch-Gordan decomposition of the tensor product of a  fully symmetric $N$-particle irreducible representation of $\rmu(D)$ times its conjugated (see later on Section \ref{Youngsec} for more details). The  coefficients $\tilde{c}_{N,n}$ will be of fundamental importance in the definition of a one-parameter $s$ family of quasi-probability distribution functions for symmetric quDits. We shall also proof  in Appendix \ref{apptaulambda} that $\tilde{c}_{N,n}$ can be written as a function of Laplacian eigenvalues $\lambda_{n}=n(n+D-1)$, a fact that will allow us to replace   $\tilde{c}_{N,n}$  by functions of the Laplacian operator $\Delta_\zb$ when acting on harmonic functions $\Psi^{(n)}_j$, thus providing differential expressions for star products and a heat kernel relation between different quasi-probability distributions (see later on Sec. \ref{heatsec}).

The orthonormal basis harmonic functions $\{\Psi^{(N)}_j, j=1,\dots,\tilde{d}_{N}\} $ of the Hilbert space $\tilde{\mathcal{H}}_N$ can be constructed as derivatives of the $\lambda_{N}$-kernel $K_N(\zb,\wb)$ as follows. Firstly compute 
\be
\tilde{\Psi}^{(N)}_{\bm{k},\bm{l}}(\zb)=\left.\frac{\partial^{k+l}K_N(\zb,\wb)}{\partial^{k_1}w_1\dots \partial^{k_{D-1}}w_{D-1}\partial^{l_1}\bar{w}_1\dots \partial^{l_{D-1}}\bar{w}_{D-1}}\right|_{\wb=\bm{0}},
\ee
where we use the shorthand $\bm{k}\equiv(k_1,\dots,k_{D-1})$ and $k\equiv k_1+\dots+k_{D-1}$ (the same for $\bm{l}$). A basis of  $\tilde{\mathcal{H}}_N$ is given by
\be
\tilde{B}_N=\left\{ \tilde{\Psi}^{(N)}_{\bm{k},\bm{l}}(\zb):  \,k\leq N,l=N, \; \mathrm{or} \; k=N, l\leq N\right\}.
\ee
Indeed, there are exactly $\tilde{d}_{N}$ independent functions $\tilde{\Psi}^{(N)}_{\bm{k},\bm{l}}$ since, for fixed $\bm{l}$ and $k\leq N$ there are $d_{N}$ functions and, therefore,  there are  $d_{N}^2$ functions when $l\leq N$ and $k\leq N$; by substracting the cases $l=N$ or $k=N$, we finally arrive to the dimension \eqref{dimtilde}. $\tilde{B}_N$ is not orthonormal in general, but we can orthonormalize it by means of the symmetric and real Gram matrix
\be
G_{\bm{k}\bm{l};\bm{k}'\bm{l}'}=\int_{\mathbb{C}^{D-1}} \tilde{\Psi}^{(N)}_{\bm{k},\bm{l}}(\bar{\zb}) \tilde{\Psi}^{(N)}_{\bm{k}',\bm{l}'}(\zb) \, d\mu_0(\zb).
\ee
We shall denote the orthonormal basis harmonic functions simply by $\Psi^{(N)}_j, j=1,\dots, \tilde{d}_{N}$.

\section{Generalized Fano multipole operators and Stratonovich-Weyl kernel}\label{FanoSWsec}

Following \cite{Brif_1998,PhysRevA.59.971} we define the $\rmu(D)$ version of the well known $\rmu(2)$ multipole operators as
\be
\hat{\mathcal{D}}^{(N)}_{n,j}\equiv \tau_{N,n}^{-1/2}\int_{\mathbb{C}^{D-1}} Y^{(n)}_j(\zb)|N,\zb\ra\la N,\zb|\,d\mu_N(\zb), \; n=0,1,\dots,N, \; j=1,\dots,\tilde{d}_{n},
\ee
where $\tau_{N,n}\equiv {d}_{N}\tilde{c}_{N,n}$ is a rescaling of $\tilde{c}_{N,n}$ in \eqref{tildec}. We are using the measure $d\mu_N(\zb)= {d}_{N}d\mu_0(\zb)$ [a rescaling of the Haar measure $d\mu_0(\zb)$ in \eqref{dmu0}] that makes the CSs $|N,\zb\ra$ to fulfill the closure relation \eqref{resolution}. This rescaling also affects the Laplacian eigenfunctions as $Y^{(n)}_j(\zb)\equiv\Psi^{(n)}_j(\zb)/\sqrt{d_{N}}$. In the Fock basis \eqref{symmetricbasis}, Fano operators are square matrices of size ${d}_{N}$ with matrix elements
\be
\la \vec{n}|\hat{\mathcal{D}}^{(N)}_{n,j}|\vec{n}'\ra=\frac{1}{\sqrt{\tau_{N,n}}}\int_{\mathbb{C}^{D-1}}  Y^{(n)}_j(\zb)\la\vec{n}|N,\zb\ra\la N,\zb|\vec{n}'\ra\,d\mu_N(\zb)
\ee
where $$\la\vec{n}|N,\zb\ra=\binom{N}{n_0,\dots,n_{D-1}}^{1/2}\frac{\prod_{i=1}^{D-1}z_i^{n_i}}{(1+\zb^\dag\zb)^{N/2}}$$
involves multinomial coefficients. 
There are 
\be \sum_{n=0}^N  \tilde{d}_{n}=d_{N}^2\label{dimyoung}\ee
generalized Fano multipole operators. From the expansion \eqref{kernelexpansion} and properties of the $\lambda_{n}$-kernel $K_n(\zb,\wb)$ one can derive the orthonormality conditions 
\be
\tr\left(\hat{\mathcal{D}}^{(N)}_{n,j}\hat{\mathcal{D}}^{(N)\dag}_{n',j'}\right)=\delta_{n,n'}\delta_{j,j'}.
\ee
Any operator $\hat{A}$ on $\mathcal{H}_N$ can be expanded as
\be
\hat{A}=\sum_{n=0}^N\sum_{j=1}^{\tilde{d}_{n}}\tr(\hat{A}\hat{\mathcal{D}}^{(N)\dag}_{n,j})\hat{\mathcal{D}}^{(N)}_{n,j}.
\ee
In particular, the Stratonovich-Weyl (SW) kernel is defined as 
\be
\hat{\Delta}^{(s)}(\zb)\equiv \sum_{n=0}^N \tau_{N,n}^{-s/2}\sum_{j=1}^{\tilde{d}_{n}} Y^{(n)}_j(\bar{\zb})\hat{\mathcal{D}}^{(N)}_{n,j}.\label{SWk}
\ee
It generalizes the Agarwal kernel for spin $j=N/2$ \cite{AgarwalPhysRevA.24.2889} from $D=2$ to arbitrary $D$. The SW kernel satisfies the standarization and tracing conditions
\be
\int_{\mathbb{C}^{D-1}}  d\mu_N(\zb)\hat{\Delta}^{(s)}_{\vec{n},\vec{n}'}(\zb)=\delta_{\vec{n},\vec{n}'},\quad \int_{\mathbb{C}^{D-1}}  d\mu_N(\zb)\hat{\Delta}^{(s)}_{\vec{n},\vec{n}'}(\zb)\hat{\Delta}^{(-s)}_{\vec{m},\vec{m}'}(\zb)=\delta_{\vec{n},\vec{m}'}\delta_{\vec{n}',\vec{m}}.\label{tracingcond}
\ee 
Any density matrix $\rho$ of $N$-symmetric quDits can be mapped onto a $s$-parameter  family of quasi-distribution functions 
\be
\mathcal{F}_\rho^{(s)}(\zb)=\tr(\rho \,\hat{\Delta}^{(s)}(\zb)),\label{sfunction}
\ee
over the  phase space ${\mathbb{C}P^{D-1}}\ni\zb$, and viceversa
\be
\rho=\int_{\mathbb{C}^{D-1}}  d\mu_N(\zb)\mathcal{F}_\rho^{(s)}(\zb)\hat{\Delta}^{(-s)}(\zb).\label{sfunctioninv}
\ee
The particular values $s=+1, 0,-1$, 
correspond to the traditional Glauber-Sudarshan $\mathcal{P}_\rho=\mathcal{F}_\rho^{(1)}$, Wigner  $\mathcal{W}_\rho=\mathcal{F}_\rho^{(0)}$, and Husimi $\mathcal{Q}_\rho=\mathcal{F}_\rho^{(-1)}$ functions, respectively.

\subsection{Case $D=2$: Wigner function for symmetric $N$-qubit systems: even and odd cat states}

Our results agree with those of  \cite{Brif_1998,PhysRevA.59.971}  for $D=2$. In particular,  the invariant  coefficients $\tau_{N,n}$ in the SW kernel \eqref{SWk} can be written in terms of SU(2) Clebsch-Gordan coefficients $\la j,j;n,0|j,j\ra$, for angular momentum $j=N/2$, as
\be
\tau_{N,n}=\frac{(N+1)}{4\pi}|\la N/2,N/2;n,0|N/2,N/2\ra|^2.\label{tau}
\ee
Reference \cite{PhysRevA.59.971} uses the familiar spherical harmonics  $Y_{n,m}(\theta,\phi), m=-n,\dots,n$ as Laplacian eigenfunctions on the sphere 
$\mathbb{S}^2$, which can be (non trivially) related to our  Laplacian eigenfunctions $Y^{(n)}_l(z), l=1,\dots,2n+1$ on the complex projective space $\mathbb{C}P^1$ parametrized by the   sthereographic projection $z=\tan(\theta/2)e^{\ic \phi}$  of a point $(\theta,\phi)$ of the Bloch sphere onto the complex plane $\mathbb{C}\ni z$.  Our invariant coefficients  \eqref{tau} differ from that of Ref.  \cite{PhysRevA.59.971}  by a factor $(N+1)/(4\pi)$  because we also  use  a different normalization, namely  $\int_{\mathbb{C}}d\mu_0(\zb)=1$, for the invariant  integration measure \eqref{dmu0}. As a particular example, the SW kernel for $N=1$ qubit is a size  $d_N=(N+1)$ square matrix adopting the simple form
\be
\hat{\Delta}^{(s)}(z)=\frac{1}{1+|z|^2}
\left(
\begin{array}{cc}
 \frac{1}{2} \left(|z|^2+1-3^{\frac{s+1}{2}} (|z|^2-1)\right) & 3^{\frac{s+1}{2}} \bar{z} \\
 3^{\frac{s+1}{2}} z & \frac{1}{2} \left(|z|^2+1+3^{\frac{s+1}{2}} (|z|^2-1)\right) \\
\end{array}
\right)
\ee
in the Fock basis \eqref{symmetricbasis}, ordered as $\{|n_0,n_1\ra, n_0+n_1=N=1\}=\{|1,0\ra,|0,1\ra\}$.

\begin{figure}
	\begin{center}
	\includegraphics[width=4.5cm]{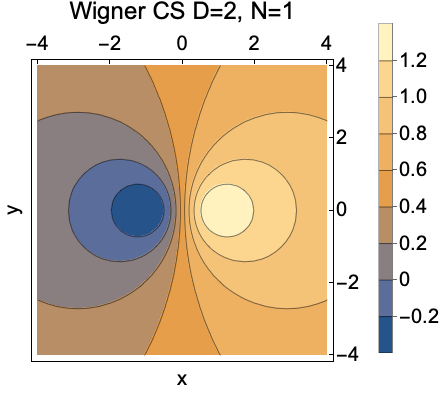}\includegraphics[width=4.5cm]{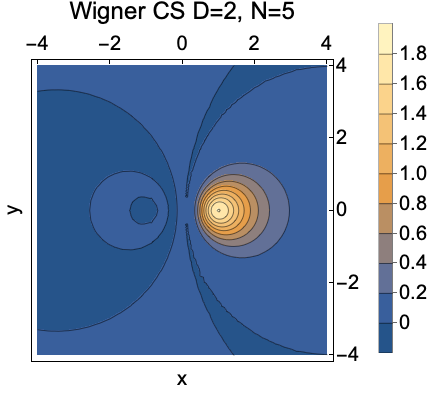}	\end{center}
	\caption{Wigner functions of spin-$N/2$ CSs $|N,z=1\ra$ of a symmetric $N$-qubit system, for $N=1$ and $N=5$ particles,  as a function of the phase-space coordinates $x_1+\ic y_1=z_1$ and $x_2+\ic y_2=z_2$. Negative value regions  end up disappearing for large $N$. }
	\label{Fig:WignerCSD2}
\end{figure}

Negativity of the Standard (harmonic oscillator) Wigner function is used as an indicator of non-classicality (see e.g. \cite{AnatoleKenfack_2004} and references therein). However, for multiqubit systems, this is not strictly true \cite{PhysRevResearch.3.033134}; for example, in Fig. \ref{Fig:WignerCSD2} we see that the Wigner function attains negative values even for coherent (quasi-classical) states $|N,z\ra$, although negative value regions  end up disappearing at the thermodynamic limit $N\to\infty$.

\begin{figure}
	\begin{center}
	\includegraphics[width=4.5cm]{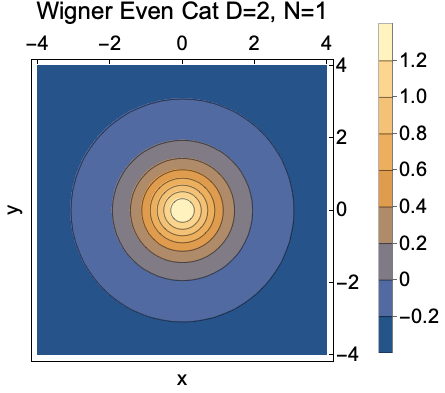}\includegraphics[width=4.5cm]{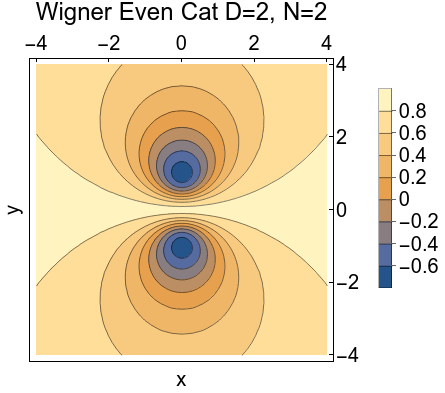}\includegraphics[width=4.5cm]{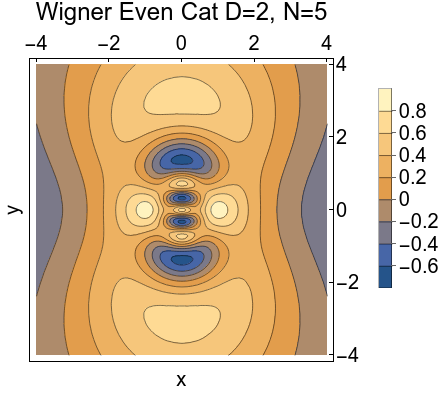}\\
	\includegraphics[width=4.5cm]{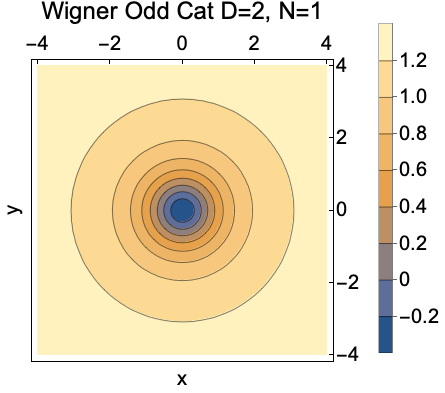}
		\includegraphics[width=4.5cm]{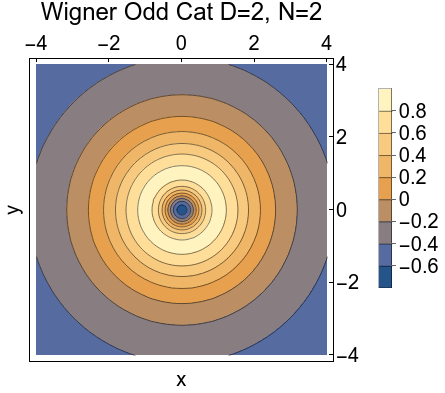}
		\includegraphics[width=4.5cm]{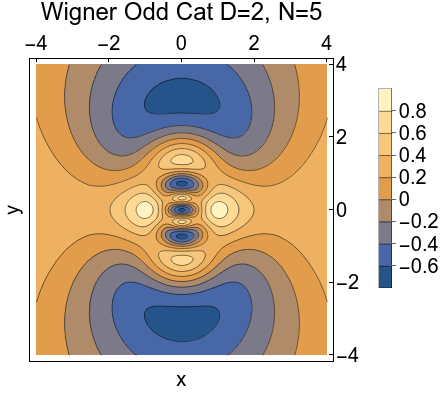}
	\end{center}
	\caption{Wigner functions of even (top panel) and odd (bottom panel) cat states $|N,1\ra_\pm$ in Eq. \eqref{cat2} of a symmetric $N$-qubit system, for $N=1,2,5$,  as a function of the phase-space coordinates $x+\ic y=z$. The values of $\mathcal{W}_{\rho_\pm}$ are arranged  roughly symmetrically about zero, especially for large $N$.}
	\label{Fig:WignerCatD2}
\end{figure}

Spin $j=N/2$ SU(2) CSs $|N,z\ra$ in \eqref{coh2} are quasi-classical states, but the even and odd (Schr\"odinger cat) normalized supperpositions 
\be
|N,z\ra_\pm=\frac{|N,z\ra\pm|N,-z\ra}{\mathcal{N}_\pm(z)}, \quad \mathcal{N}_\pm(z)=\sqrt{2}\left[1\pm \left(\frac{1-|z|^2}{1+|z|^2}\right)^N\right]^{1/2},\label{cat2}
\ee
display non-classical properties as: non-Poissonian statistics, squeezing, entanglement, etc. (see e.g. \cite{Calixto_2017} for multi-qubit and \cite{QIP-2021-Entanglement} for multi-quDit systems). Note that, for $z=1$, the CSs $|N,z\ra$ and $|N,-z\ra$ are orthogonal and, therefore, even and odd cats $|N,1\ra_\pm$ are Bell-like states. In Figure \ref{Fig:WignerCatD2}  we plot the Wigner functions $\mathcal{W}_{\rho_\pm}(x,y)$ of the density matrices 
$\rho_\pm=|N,1\ra_\pm\la N,1|$ of these particular even and odd cat (Bell-like) states as a function of the phase-space coordinates $x+\ic y=z$. As an example, the explicit  expression for $N=2$ qubits is
\bea
\mathcal{W}_{\rho_+}(x,y)&=&\frac{1}{6} \left(\sqrt{10}+2\right)-\frac{2 \sqrt{10} y^2}{\left(x^2+y^2+1\right)^2},\\
\mathcal{W}_{\rho_-}(x,y)&=&\frac{2 \sqrt{10}}{x^2+y^2+1}-\frac{2 \sqrt{10}}{\left(x^2+y^2+1\right)^2}-\frac{\sqrt{10}}{3}+\frac{1}{3}.
\eea
Dark colors of the contourplots  Figure \ref{Fig:WignerCatD2} indicate those phase-space regions where cat states are highly non-classical (negative Wigner function), specially for higher values of $N$.

\subsection{Case $D=3$: Wigner function for symmetric $N$-qutrit systems: even and odd cat states}

As a particular example, the SW kernel for $N=1$ qutrit is now a size $d_1=(N+1)(N+2)/2=3$ matrix adopting the simple form

\be
\hat{\Delta}^{(s)}(\zb)=\frac{2^{1+s}}{1+|z_1|^2+|z_2|^2}\left(
\begin{array}{ccc}
\frac{2^{-s}+4+(2^{-s}-2)(|z_1|^2+|z_2|^2)}{6}& \bar{z}_1 & \bar{z}_2 \\
 z_1 & \frac{(2^{-s}+4)|z_1|^2+(2^{-s}-2)(1+|z_2|^2)}{6}& z_1 \bar{z}_2 \\
 z_2 & z_2 \bar{z}_1 & \frac{(2^{-s}+4)|z_2|^2+(2^{-s}-2)(1+|z_1|^2)}{6}
\end{array}
\right)
\ee

in the Fock basis \eqref{symmetricbasis}, ordered as $\{|n_0,n_1,n_2\ra, n_0+n_1+n_2=N\}=\{|1,0,0\ra,|0,1,0\ra,|0,0,1\ra\}$.

\begin{figure}
	\begin{center}
	\includegraphics[width=4.5cm]{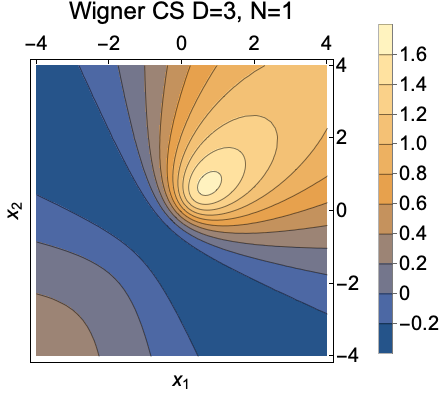}\includegraphics[width=4.5cm]{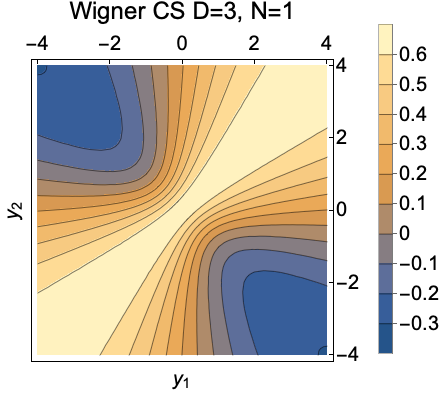}\includegraphics[width=4.5cm]{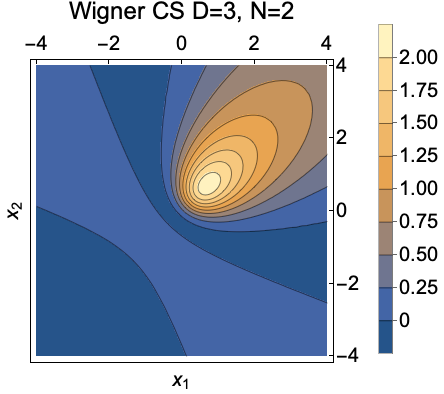}\includegraphics[width=4.5cm]{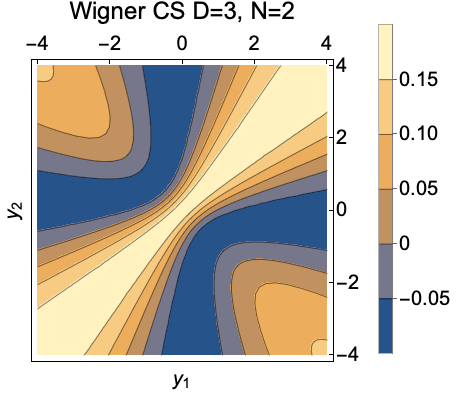}	\end{center}
	\caption{Wigner functions of U(3)-spin CSs $|N,1/\sqrt{2},1/\sqrt{2}\ra$ of a symmetric $N$-qutrit system, for $N=1$ and $N=2$ particles,  as a function of the phase-space coordinates $x_1+\ic y_1=z_1$ and $x_2+\ic y_2=z_2$.}
	\label{Fig:WignerCSD3}
\end{figure}

\begin{figure}
	\begin{center}
	\includegraphics[width=4.5cm]{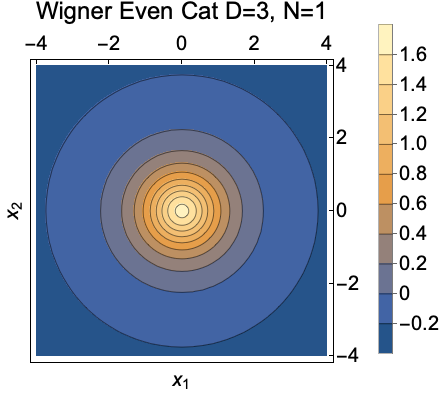}\includegraphics[width=4.5cm]{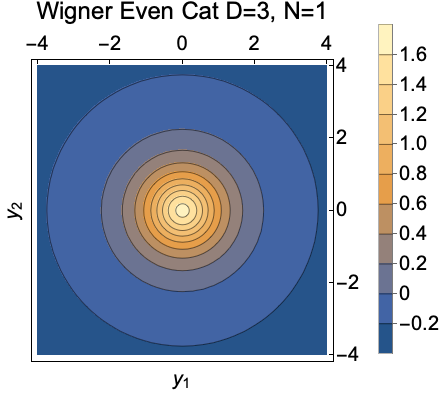}\includegraphics[width=4.5cm]{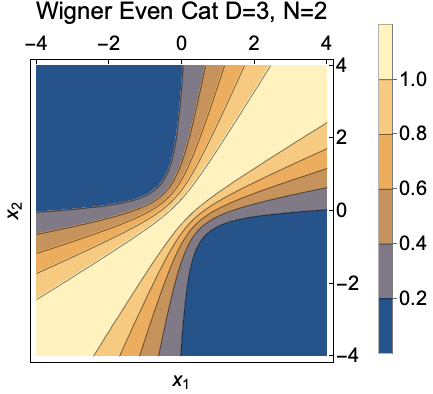}\includegraphics[width=4.5cm]{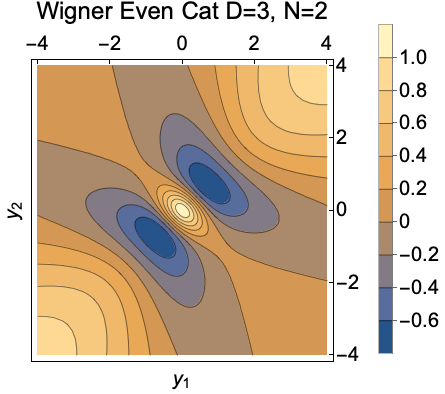}\\
	\includegraphics[width=4.5cm]{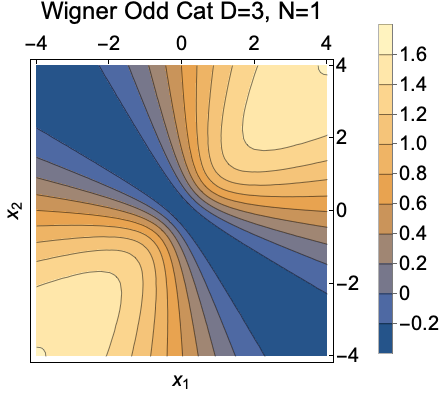}\includegraphics[width=4.5cm]{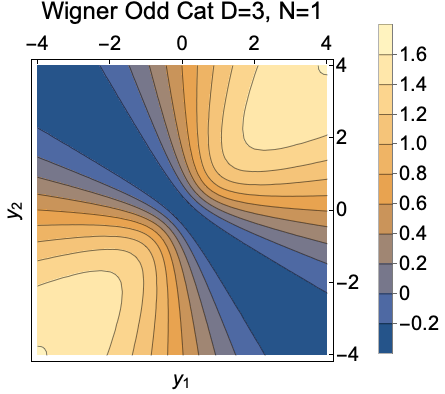}\includegraphics[width=4.5cm]{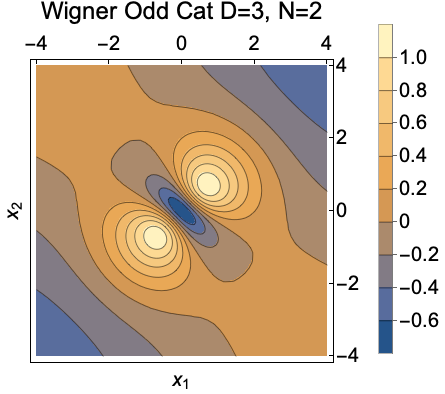}\includegraphics[width=4.5cm]{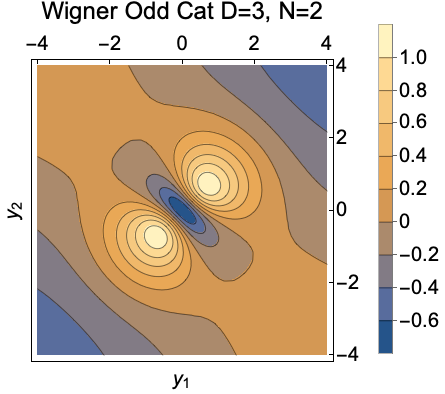}
	\end{center}
	\caption{Wigner functions of totally even ($\mathbbm{c}=[0,0]$, top panel) and odd ($\mathbbm{c}=[1,1]$, bottom panel) cat states $|N,1/\sqrt{2},1/\sqrt{2}\ra_\mathbbm{c}$ in Eq. \eqref{S3C} of a symmetric $N$-qutrit system, for $N=1$ and $N=2$ particles,  as a function of the phase-space coordinates $x_1+\ic y_1=z_1$ and $x_2+\ic y_2=z_2$.}
	\label{Fig:WignerCatD3}
\end{figure}

Now we want to analyze the phase space structure of $N$-qutrit Schr\"odinger cats as parity adapted U$(3)$-spin CSs. We address the reader to   \cite{nuestroPRE,QIP-2021-Entanglement,Guerrero22,Guerrero_2023,Mayorgas2023} for a discussion of the general $D$ case. 
For $D=3$ we have $2^{D-1}=4$ parity sectors (the elements of the parity $\mathbb{Z}_2^{D-1}$ dual group) denoted as 
\be
\mathbbm{c}=[c_1,c_2]\in\big\{[0,0], \, [0,1], \, [1,0], \, [1,1]\big\},
\ee
and therefore four Schr\"odinger cat states associated to the four different parity adaptations of the CS $|N,\zb\ra=|N,z_1,z_2\ra$ defined in \eqref{cohD} , adopting the explicit form
\begin{align}\label{S3C}
|N,\zb\ra_{\mathbbm{c}}=\frac{1}{4\mathcal{N}_\mathbbm{c}(\zb)}\Big[&\,|N,z_1,z_2\ra
	+(-1)^{c_1}|N,-z_1,z_2\ra+(-1)^{c_2}|N,z_1,-z_2\ra
	+(-1)^{c_1+c_2}|N,-z_1,-z_2\ra\Big],
\end{align}
with squared norm 
\begin{align}\label{S3CN}
	\mathcal{N}_{\mathbbm{c}}(\zb)^2=&\,\frac{(1+|z_1|^2+|z_2|^2)^N+(-1)^{c_1} (1-|z_1|^2+|z_2|^2)^N+(-1)^{c_2} (1+|z_1|^2-|z_2|^2)^N+(-1)^{c_1+c_2} (1-|z_1|^2-|z_2|^2)^N}{4 (1+|z_1|^2+|z_2|^2)^N}.
\end{align}
Let us pick up the multimode even $(+)$  and odd $(-)$ cat combinations \cite{Manko-MultimodeCats}
\be
|N,\zb\ra_+=\frac{1}{\sqrt{2}}\left(|N,\zb\ra_{[0,0]}+|N,\zb\ra_{[1,1]}\right),\quad |N,\zb\ra_-=\frac{1}{\sqrt{2}}\left(|N,\zb\ra_{[0,1]}+|N,\zb\ra_{[1,0]}\right).
\ee
In Figure \ref{Fig:WignerCatD3}  we plot the Wigner functions $\mathcal{W}_{\rho_\pm}(\zb)$ of the density matrices 
$$\rho_\pm=\Big|N,\frac{1}{\sqrt{2}},\frac{1}{\sqrt{2}}\Big\ra_\pm\Big\la N,\frac{1}{\sqrt{2}},\frac{1}{\sqrt{2}}\Big|$$ of these particular totally even and odd cat (Bell-like) states as a function of the phase-space coordinates $x_1+\ic y_1=z_1$ and $x_2+\ic y_2=z_2$. We show section plots in ``position" $(x_1,x_2)$ and ``momentum" $(y_1,y_2)$ spaces. Dark colors of the contourplots  Figure \ref{Fig:WignerCatD2} indicate those phase-space regions where cat states are highly non-classical (negative Wigner function), specially for higher values of $N$. However, as mentioned above, our Wigner function can also attain negative values for supposedly quasi-classical states such as coherent states $|N,\zb\ra$. Indeed, Figure \ref{Fig:WignerCSD3} shows negative value regions for $\rmu(3)$-spin coherent states, although it is true that the Wigner function is  predominantly positive in this case, especially for larger values of $N$.

\section{Young tableaux and Clebsch-Gordan decomposition of Kronecker products}\label{Youngsec}

Let us provide an interesting  diagramatic picture of our construction in terms of Young tableaux associated to an irreducible representation of U$(D)$. In the standard notation, a Young tableau of shape $h=[h_1,\dots,h_D]$ is represented by
\begin{equation}\label{youngdiagram}
	\overbrace{
		\begin{gathered}
			\begin{ytableau}
				~ &...&...&...&...&...&~\\
				: &:& : & : &:\\
				~&...&~
			\end{ytableau}
		\end{gathered}
	}^{h_1}
\end{equation}
that is, a diagram  of  $N=h_1+\dots h_D$ boxes/particles ($h$ is also called a partition of $N$)
with  $h_1\geq \dots \geq h_D$ and $h_i$ the number of boxes in row $i=1,\dots,D$. One  sometimes uses the shorthand 
$h=[n,\stackrel{M}{\dots},n,0,\dots ,0]=[n^M]$ to write Young diagrams with $M$ rows of equal length $n$, obviating zero-box rows. The dimension formula for a general irreducible representation $h=[h_1,\dots,h_D]$ of U$(D)$ is given by
\begin{equation}
 d[h]=\frac{\prod_{i<j}(h_i-h_j+j-i)}{\prod_{i=1}^{D-1}i!},
\end{equation}
which comes from the Hook length formula \cite{Barut}.  

The expansion \eqref{CGexpansion} of the  $N$-particle squared norm CS overlap  $Q^N(\zb,\wb)$ in \eqref{csoverlapN} in terms of $\lambda_{n}$-kernels  $K_n(\zb,\wb)$, can be related to the Clebsch-Gordan decomposition 
\be
 \stackrel{d_N}{[N]}\otimes  \stackrel{d_N}{[N^{D-1}]}= \bigoplus_{n=0}^N  \stackrel{ \tilde{d}_{n}}{[N+n,N^{D-2},N-n]} \quad \longleftrightarrow \quad d_{N}^2=\sum_{n=0}^N  \tilde{d}_{n}
 \ee
of the Kronecker product $[N]\otimes [N^{D-1}]$ of the fully symmetric representation $[N]$ of $N$ quDits times its conjugated representation $ [N^{D-1}]$ (both with dimension $d_{N}=\tbinom{N+D-1}{N}$)  into irreducibles $[N+n,N^{D-2},N-n]$. We are writting the corresponding dimension of the irreducible representation on top of each Young diagram. 
Indeed, one can realize that the dimension $d[N+n,N^{D-2},N-n]$ coincides with the dimension  $\tilde{d}_{n}$  in \eqref{dimtilde}  of the Hilbert space $\tilde{\mathcal{H}}_n$ of $\lambda_n$-eigenfunctions of the Laplacian. Therefore, the dimensions of this Clebsch-Gordan decomposition are in accordance with the formula $d_{N}^2=\sum_{n=0}^N  \tilde{d}_{n}$ already reported in Eq. \eqref{dimyoung}. For example, for $N=2$ qutrits we have

\be \Yvcentermath1 \stackrel{6}{\yng(2)}\otimes \stackrel{6}{\yng(2,2)} = 
\stackrel{1}{\yng(2,2,2)}\oplus  \stackrel{8}{\yng(3,2,1)} \oplus \stackrel{27}{\yng(4,2)}\label{octete}. \ee
which indeed gives $6^2=1+8+27$.

\section{Heat kernel and Gaussian-like smoothing of quasi-probability distributions}\label{heatsec}

Now we investigate on an interesting connection with the heat equation $\frac{\partial\Phi}{\partial t}=\Delta_{\alphab}\Phi$ and the relation between the $\lambda_N$-kernel \eqref{kernelexpansion}  and a generalized heat kernel.

Inserting the tracing condition \eqref{tracingcond} into the expression \eqref{sfunction} we get a relationship between quasi-probability distribution  functions 

\be
\mathcal{F}_\rho^{(s)}(\zb)=\int_{\mathbb{C}^{D-1}}  d\mu_N(\zb')\mathcal{K}_{s,s'}(\zb,\zb')\mathcal{F}_\rho^{(s')}(\zb'),\quad \mathcal{K}_{s,s'}(\zb,\zb')=\tr[\hat{\Delta}^{(s)}(\zb)\hat{\Delta}^{(-s')}(\zb')].\label{integralHK}
\ee
In particular, if we take the case $s=-1$ and $s'=0$, we get a relationship between the Husimi and the Wigner functions 
\be
\mathcal{Q}_\rho(\zb)=\int_{\mathbb{C}^{D-1}}  d\mu_N(\zb')\mathcal{K}_{-1,0}(\zb,\zb')\mathcal{W}_\rho(\zb').
\ee

For the flat/Euclidean (harmonic oscillator) case, the kernel $\mathcal{K}_{-1,0}$ has been related in the literature (see e.g. \cite{Vourdas_2006})   to the ``heat kernel'' $\mathcal{K}_{-1,0}=\exp(t_0 \Delta_{\alphab})$ of the standard heat equation $\frac{\partial\Phi}{\partial t}=\Delta_{\alphab}\Phi$, where $t_0$ is the ``Gaussian smoothing time'' we need to evolve $\mathcal{W}_\rho$ to arrive to $\mathcal{Q}_\rho$. 

In our non-flat/complex-projective (symmetric $N$-quDit) case, we can arrive to a deformed heat equation with ``time'' parameter $t=s$, which tends to the standard heat equation in the thermodynamic (large number of particles) limit $N\to\infty$. Indeed, in Appendix \ref{apptaulambda} we show how to write the invariant (generalized Clebsch-Gordan) coefficients $\tau_{N,n}$ in the SW kernel \eqref{SWk} as  a function $\tau_{N,n}=T_N(\lambda_n)$ of the Laplacian $\Delta_{\zb}$ eigenvalues $\lambda_n=n(n+D-1)$. Therefore, when multiplying Laplacian eigenfunctions, we can formally write
\be
\tau_{N,n} Y^{(n)}_j(\zb)=T_N(\Delta_{\zb})Y^{(n)}_j(\zb).\label{TDelta}
\ee
Hence, deriving the SW kernel \eqref{SWk} with respect to the parameter $s$ we get
\be
\frac{\partial\hat{\Delta}^{(s)}(\zb)}{\partial s}=-\frac{1}{2}\ln[T_N(\Delta_{\zb})]\hat{\Delta}^{(s)}(\zb),\label{SWevol}
\ee
which is the announced generalization of the standard heat equation $\frac{\partial\Phi}{\partial t}=\Delta_{\alphab}\Phi$, with $s$ playing the role of ``time" $t$. Actually, in the large number of particles limit $N\gg 1$, the expansion \eqref{lambdaexpansion} gives 
\be
\frac{\partial\hat{\Delta}^{(s)}(\zb)}{\partial s}\simeq \frac{1}{2N}\Delta_{\zb}\hat{\Delta}^{(s)}(\zb),
\ee
which, after a necessary rescaling $\zb=\alphab/\sqrt{N}$ and $t=s/2$,  recovers the standard heat equation with Laplacian 
\be
\Delta_{\alphab}=\lim_{N\to\infty}\frac{1}{N}\Delta_{\zb}=\sum_{i=1}^{D-1}\frac{\partial^2}{\partial \bar{\alpha}_i\partial \alpha_i}.
\ee
For $D=2$, this large $N$ limit is related to the usual Holstein-Primakoff \cite{HolsteinPrimakoff} approximation mapping  atomic/spin  CSs $|\zb\ra$ onto harmonic oscillator CSs $|\alphab\ra$ (see also  \cite{JMRadcliffe_1971,GilmorePhysRevA.6.2211}), in turn related to a group contraction from the SU(2) angular momentum algebra to the harmonic oscillator algebra.

The  kernel $\mathcal{K}_{s,s'}(\zb,\zb')$ connecting two quasi-probability distribution functions $\mathcal{F}_\rho^{(s)}(\zb)$ and $\mathcal{F}_\rho^{(s')}(\zb')$ in \eqref{integralHK}  can then be written as a generalized heat kernel 

\be
\mathcal{K}_{s,s'}(\zb,\zb')=\tr\left[\hat{\Delta}^{(0)}(\zb')[T_N(\Delta_{\zb})]^{(s'-s)/2}\hat{\Delta}^{(0)}(\zb)\right]=\frac{1}{d_N}\sum_{n=0}^N \tau_{N,n}^{(s'-s)/2}K_n(\zb,\zb'),\label{heatkernel}
\ee
 in terms of the $\lambda_n$-kernel \eqref{kernelexpansion}, where we have used \eqref{SWevol} and the tracing conditions \eqref{tracingcond}.
Therefore, in general, $\mathcal{F}_\rho^{(s)}(\zb)$ can be seen as a (generalized) ``Gaussian smoothing'' of $\mathcal{F}_\rho^{(s')}(\zb)$, for a smoothing time $t=(s'-s)/2$, with $s'>s$ and generalized Gaussian heat kernel $\mathcal{K}_{s,s'}(\zb,\zb')$. In particular, the Husimi function $\mathcal{Q}_\rho=\mathcal{F}_\rho^{(-1)}$ is a Gaussian-like smoothing of the Glauber-Sudarshan $\mathcal{P}_\rho=\mathcal{F}_\rho^{(1)}$ and Wigner  $\mathcal{W}_\rho=\mathcal{F}_\rho^{(0)}$ functions with smoothing times $t=(s'-s)/2=1$ and $t=1/2$, respectively. Actually, the heat hernel 
$\mathcal{K}_{-1,1}(\zb,\zb')$ connecting $\mathcal{P}_\rho$ and $\mathcal{Q}_\rho$ functions coincides with the CS overlap $Q^N(\zb,\wb)$ according to \eqref{CGexpansion}.

\section{Generalized twisted product and Moyal bracket}\label{twistsec}

To finish, we compute the twisted (Moyal, star) product 
\be
\mathcal{F}_{\hat{A}}^{(s')}(\zb)  \star \mathcal{F}_{\hat{B}}^{(s'')}(\zb)\equiv  \mathcal{F}_{\hat{A}\hat{B}}^{(s)}(\zb) 
\ee
of phase-space functions $\mathcal{F}_{\hat{A}}^{(s')}(\zb)$ and  $\mathcal{F}_{\hat{B}}^{(s'')}(\zb)$ corresponding to the usual  product of operators $\hat{A}\hat{B}$. Using \eqref{sfunction} and \eqref{sfunctioninv}, we have
\be
 \mathcal{F}_{\hat{A}\hat{B}}^{(s)}(\zb)=\tr(\hat{\Delta}^{(s)}(\zb)\,\hat{A}\hat{B})=\int_{\mathbb{C}^{D-1}}  d\mu_N(\zb')\int_{\mathbb{C}^{D-1}}  d\mu_N(\zb'')\mathcal{L}^{s}_{s',s''}(\zb,\zb',\zb'')\mathcal{F}_{\hat{A}}^{(s')}(\zb')  \mathcal{F}_{\hat{B}}^{(s'')}(\zb''),
 \ee
 with the trikernel
 \bea
 \mathcal{L}^{s}_{s',s''}(\zb,\zb',\zb'')&\equiv&\tr\left(\hat{\Delta}^{(s)}(\zb)\hat{\Delta}^{(-s')}(\zb')\hat{\Delta}^{(-s'')}(\zb'')\right)\nonumber\\ 
 &=& \tr\left( \hat{\Delta}^{(0)}(\zb)[T_N(\Delta_{\zb'})]^{(s'-s)/2} \Delta^{(0)}(\zb') [T_N(\Delta_{\zb''})]^{(s''+s')/2} \Delta^{(0)}(\zb'') \right).\label{trikernel}
 \eea
 Using the standarization condition \eqref{tracingcond}, we find the relation
 \begin{equation}
  \int_{\mathbb{C}^{D-1}}  d\mu_N(\zb)\mathcal{L}^{s}_{s',s''}(\zb,\zb',\zb'')=\mathcal{K}_{-s',s''}(\zb',\zb'')
 \end{equation}
between the trikernel and the generalized heat (bi-)kernel \eqref{heatkernel}.

The Moyal bracket is then defined for  Wigner functions as 
\begin{equation}
 [\mathcal{F}_{\hat{A}}^{(0)}(\zb), \mathcal{F}_{\hat{B}}^{(0)}(\zb)]_\mathrm{M}\equiv-\ic\left(\mathcal{F}_{\hat{A}}^{(0)}(\zb)\star \mathcal{F}_{\hat{B}}^{(0)}(\zb)- \mathcal{F}_{\hat{B}}^{(0)}(\zb)\star  \mathcal{F}_{\hat{A}}^{(0)}(\zb)\right).
\end{equation}
Using the differential form of the trikernel \eqref{trikernel} in terms of the Laplacian \eqref{Laplacian}, and using the expansion \eqref{lambdaexpansion} one finds the Poisson bracket
\begin{equation}
\left\{\mathcal{F}_{\hat{A}}^{(0)}(\zb),\mathcal{F}_{\hat{B}}^{(0)}(\zb)\right\}\equiv \ic(1+\zb^\dag\zb)\sum_{i,j=1}^{D-1}(\delta_{ij}+\bar{z}_iz_j)\left(\frac{\partial \mathcal{F}_{\hat{A}}^{(0)}(\zb)}{\partial \bar{z}_i}\frac{\partial \mathcal{F}_{\hat{B}}^{(0)}(\zb)}{\partial z_j}-\frac{\partial \mathcal{F}_{\hat{B}}^{(0)}(\zb)}{\partial \bar{z}_i}\frac{\partial \mathcal{F}_{\hat{A}}^{(0)}(\zb)}{\partial z_j}\right),
\end{equation}
as the leading term in the $1/N$ expansion of the Moyal bracket. The thermodynamical limit $N\to\infty$ plays here the same role as the usual  $\hbar\to 0$ in semiclassical expansions. 
For the particular case of $D=2$, and using the stereographic projection $z=\tan(\theta/2)e^{\ic \phi}$,  we recover the usual Poisson bracket 
\begin{equation}
 \left\{f(\theta,\phi),g(\theta,\phi)\right\}=\frac{1}{\sin(\theta)}\left(\frac{\partial f(\theta,\phi)}{\partial\theta}\frac{\partial g(\theta,\phi)}{\partial\phi}-\frac{\partial g(\theta,\phi)}{\partial\theta}\frac{\partial f(\theta,\phi)}{\partial\phi}\right),
\end{equation}
for functions $f,g$ on the sphere.

\section{Conclusions and outlook}\label{conclusec}

Extending the Stratonovich-Weyl construction from Heisenberg-Weyl symmetries (and harmonic oscillators) to arbitrary symmetry groups $G$ is not a trivial task. In this article we have tackled the case $G=\rmu(D)$, restricting ourselves to fully symmetric representations (symmetric $N$-quDit systems) with  phase spaces the complex projective $\mathbb{C}P^{D-1}=\rmu(D)/\rmu(D-1)\times \rmu(1)$. We have found eigenfunctions $\Psi^{(N)}_j$ of the Laplace-Beltrami operator $\Delta_\zb$ through  coherent state overlaps $Q^N$ and the $\lambda_N$-kernel $K_N$.  Expanding $Q^N$ in terms of $K_n, n=0,\dots,N$ provides essential invariant coefficients $\tilde{\tau}_{N,n}$ to define generalized Fano multipole operators $\hat{\mathcal{D}}^{(N)}_{n,j}$ and the Stratonovich-Weyl kernel $\hat{\Delta}^{(s)}$, from which a family of quasi-probability distributions $\mathcal{F}^{(s)}_\rho$ can be constructed for a density matrix $\rho$. The case $s=0$ (Wigner function) is used to analyze the phase-space structure of parity adapted coherent states (Schrödinger cats), specially in the cases $D=2$ and $D=3$ (symmetric multi-qubits and multi-qutrits, respectively). The generalized heat kernel $\mathcal{K}_{s,s'}$ conecting two quasi-probability distribution functions $\mathcal{F}^{(s)}_\rho$ and  $\mathcal{F}^{(s')}_\rho$ has been computed as an expansion in terms of  $\lambda_n$-kernels  $K_n$ with invariant coefficients $\tilde{\tau}_{N,n}^{(s'-s)/2}$. The generalized heat bi- and tri-kernels allow for differential and integral expressions of star-products, recovering the flat case in the thermodynamic limit $N\to\infty$. A diagramatic picture of our construction  in terms of Young tableaux has also been provided. This picture could be useful when trying to extend this construction to more general irreducible representations of $\rmu(D)$ (``fermionic mixtures") other than the fully symmetric representation; this can be the physical case of fermionic alkaline-earth atomic gases trapped in optical lattices \cite{PhysRevLett.92.170403,PhysRevA.99.063605,Cazalilla_2009,PhysRevA.96.032701}. This is still work in progress.

\section*{Acknowledgments}
We thank the support of the Spanish Ministry of Science through the project PID2022-138144NB-I00.

\appendix

\section{Invariant coefficients $\tau_{N,n}$ as a function of Laplacian eigenvalues $\lambda_{n}$}\label{apptaulambda}

We are going to show that the invariant coefficients $\tau_{N,n}\equiv {d}_{N}\tilde{c}_{N,n}$ can be written as functions of Laplacian eigenvalues $\lambda_n=n(n+D-1)$. For $D=2$, the proof was done in \cite{Klimov_2002,Klimov_JMP2002}; here we generalize it to arbitrary $D$. We only need to deal with the factor $(N-n)!(N+n+D-1)!$ in \eqref{tildec} and try to write it in terms of $\lambda_n$. We shall consider  $h=1/(N+D-1)$  as small expansion ``semiclassical" parameter for a  large number $N$ of particles: First we realize that 
\be
(N-n)!(N+n+D-1)!=(N+D-2)!(N+1)!\prod_{k=0}^{n}\frac{1+kh}{1-(k+D-2)h}, \quad n=0,1,\dots,N.
\ee
 Using the expansion
 \be
 \ln \left(\frac{1+k h}{1-(k+D-2)h}\right)=\sum _{m=1}^{\infty } \left(\frac{(k+D-2)^m}{m}+\frac{(-1)^{m+1} k^m}{m}\right)h^m
 \ee 
 and that 
 \be
 \sum _{k=0}^n \frac{(k+D-2)^m}{m}+\frac{(-1)^{m+1} k^m}{m}=-\frac{\zeta (-m,n+D-1)-\zeta (-m,D-2)+(-1)^m H_n^{(-m)}}{m}
 \ee
[the $l$-th generalized harmonic number $H_n^{(l)}$ and the Hurwitz zeta function $\zeta(l,a)$] we  arrive to
\be
(N-n)!(N+n+D-1)!=(N+D-2)!(N+1)!\exp\left[(\lambda_n+D-2)\left(h+(D-2) \frac{h^2}{2}+O(h^3)\right)\right].
\ee
From here, the invariant  coefficients $\tau_{N,n}$ in the SW kernel \eqref{SWk} can be written as  a function of the Laplacian eigenvalues $\lambda_n$ as
\be
\tau_{N,n}=T_N(\lambda_n)=\exp\left[-\frac{\lambda_n }{N}+\frac{D \lambda_n }{2 N^2}-\frac{\lambda_n  (D(2 D-1)+\lambda_n )}{6 N^3}+O\left(\frac{1}{N^{7/2}}\right)\right],\label{lambdaexpansion}
\ee
in the large number of particles ($N\gg 1$) limit.

\bibliography{bibliografia.bib}

\begin{thebibliography}{63}%
\makeatletter
\providecommand \@ifxundefined [1]{%
 \@ifx{#1\undefined}
}%
\providecommand \@ifnum [1]{%
 \ifnum #1\expandafter \@firstoftwo
 \else \expandafter \@secondoftwo
 \fi
}%
\providecommand \@ifx [1]{%
 \ifx #1\expandafter \@firstoftwo
 \else \expandafter \@secondoftwo
 \fi
}%
\providecommand \natexlab [1]{#1}%
\providecommand \enquote  [1]{``#1''}%
\providecommand \bibnamefont  [1]{#1}%
\providecommand \bibfnamefont [1]{#1}%
\providecommand \citenamefont [1]{#1}%
\providecommand \href@noop [0]{\@secondoftwo}%
\providecommand \href [0]{\begingroup \@sanitize@url \@href}%
\providecommand \@href[1]{\@@startlink{#1}\@@href}%
\providecommand \@@href[1]{\endgroup#1\@@endlink}%
\providecommand \@sanitize@url [0]{\catcode `\\12\catcode `\$12\catcode
  `\&12\catcode `\#12\catcode `\^12\catcode `\_12\catcode `\%12\relax}%
\providecommand \@@startlink[1]{}%
\providecommand \@@endlink[0]{}%
\providecommand \url  [0]{\begingroup\@sanitize@url \@url }%
\providecommand \@url [1]{\endgroup\@href {#1}{\urlprefix }}%
\providecommand \urlprefix  [0]{URL }%
\providecommand \Eprint [0]{\href }%
\providecommand \doibase [0]{http://dx.doi.org/}%
\providecommand \selectlanguage [0]{\@gobble}%
\providecommand \bibinfo  [0]{\@secondoftwo}%
\providecommand \bibfield  [0]{\@secondoftwo}%
\providecommand \translation [1]{[#1]}%
\providecommand \BibitemOpen [0]{}%
\providecommand \bibitemStop [0]{}%
\providecommand \bibitemNoStop [0]{.\EOS\space}%
\providecommand \EOS [0]{\spacefactor3000\relax}%
\providecommand \BibitemShut  [1]{\csname bibitem#1\endcsname}%
\let\auto@bib@innerbib\@empty
\bibitem [{\citenamefont {Wigner}(1932)}]{PhysRev.40.749}%
  \BibitemOpen
  \bibfield  {author} {\bibinfo {author} {\bibfnamefont {E.}~\bibnamefont
  {Wigner}},\ }\bibfield  {title} {\enquote {\bibinfo {title} {On the quantum
  correction for thermodynamic equilibrium},}\ }\href {\doibase
  10.1103/PhysRev.40.749} {\bibfield  {journal} {\bibinfo  {journal} {Phys.
  Rev.}\ }\textbf {\bibinfo {volume} {40}},\ \bibinfo {pages} {749--759}
  (\bibinfo {year} {1932})}\BibitemShut {NoStop}%
\bibitem [{\citenamefont {Husimi}(1940)}]{husimi1940}%
  \BibitemOpen
  \bibfield  {author} {\bibinfo {author} {\bibfnamefont {Kôdi}\ \bibnamefont
  {Husimi}},\ }\bibfield  {title} {\enquote {\bibinfo {title} {Some formal
  properties of the density matrix},}\ }\href {\doibase
  10.11429/ppmsj1919.22.4_264} {\bibfield  {journal} {\bibinfo  {journal}
  {Proceedings of the Physico-Mathematical Society of Japan. 3rd Series}\
  }\textbf {\bibinfo {volume} {22}},\ \bibinfo {pages} {264–314} (\bibinfo
  {year} {1940})}\BibitemShut {NoStop}%
\bibitem [{\citenamefont
  {Glauber}(1963{\natexlab{a}})}]{GlauberPhysRevLett.10.84}%
  \BibitemOpen
  \bibfield  {author} {\bibinfo {author} {\bibfnamefont {Roy~J.}\ \bibnamefont
  {Glauber}},\ }\bibfield  {title} {\enquote {\bibinfo {title} {Photon
  correlations},}\ }\href {\doibase 10.1103/PhysRevLett.10.84} {\bibfield
  {journal} {\bibinfo  {journal} {Phys. Rev. Lett.}\ }\textbf {\bibinfo
  {volume} {10}},\ \bibinfo {pages} {84--86} (\bibinfo {year}
  {1963}{\natexlab{a}})}\BibitemShut {NoStop}%
\bibitem [{\citenamefont {Glauber}(1963{\natexlab{b}})}]{PhysRev.131.2766}%
  \BibitemOpen
  \bibfield  {author} {\bibinfo {author} {\bibfnamefont {Roy~J.}\ \bibnamefont
  {Glauber}},\ }\bibfield  {title} {\enquote {\bibinfo {title} {Coherent and
  incoherent states of the radiation field},}\ }\href {\doibase
  10.1103/PhysRev.131.2766} {\bibfield  {journal} {\bibinfo  {journal} {Phys.
  Rev.}\ }\textbf {\bibinfo {volume} {131}},\ \bibinfo {pages} {2766–2788}
  (\bibinfo {year} {1963}{\natexlab{b}})}\BibitemShut {NoStop}%
\bibitem [{\citenamefont {Sudarshan}(1963)}]{PhysRevLett.10.277}%
  \BibitemOpen
  \bibfield  {author} {\bibinfo {author} {\bibfnamefont {E.~C.~G.}\
  \bibnamefont {Sudarshan}},\ }\bibfield  {title} {\enquote {\bibinfo {title}
  {Equivalence of semiclassical and quantum mechanical descriptions of
  statistical light beams},}\ }\href {\doibase 10.1103/PhysRevLett.10.277}
  {\bibfield  {journal} {\bibinfo  {journal} {Phys. Rev. Lett.}\ }\textbf
  {\bibinfo {volume} {10}},\ \bibinfo {pages} {277--279} (\bibinfo {year}
  {1963})}\BibitemShut {NoStop}%
\bibitem [{\citenamefont {Weyl}(1927)}]{Weyl1927}%
  \BibitemOpen
  \bibfield  {author} {\bibinfo {author} {\bibfnamefont {H.}~\bibnamefont
  {Weyl}},\ }\bibfield  {title} {\enquote {\bibinfo {title} {Quantenmechanik
  und gruppentheorie},}\ }\href {\doibase 10.1007/BF02055756} {\bibfield
  {journal} {\bibinfo  {journal} {Zeitschrift f{\"u}r Physik}\ }\textbf
  {\bibinfo {volume} {46}},\ \bibinfo {pages} {1--46} (\bibinfo {year}
  {1927})}\BibitemShut {NoStop}%
\bibitem [{\citenamefont {Groenewold}(1946)}]{GROENEWOLD1946405}%
  \BibitemOpen
  \bibfield  {author} {\bibinfo {author} {\bibfnamefont {H.J.}\ \bibnamefont
  {Groenewold}},\ }\bibfield  {title} {\enquote {\bibinfo {title} {On the
  principles of elementary quantum mechanics},}\ }\href {\doibase
  https://doi.org/10.1016/S0031-8914(46)80059-4} {\bibfield  {journal}
  {\bibinfo  {journal} {Physica}\ }\textbf {\bibinfo {volume} {12}},\ \bibinfo
  {pages} {405--460} (\bibinfo {year} {1946})}\BibitemShut {NoStop}%
\bibitem [{\citenamefont {Moyal}(1949)}]{Moyal_1949}%
  \BibitemOpen
  \bibfield  {author} {\bibinfo {author} {\bibfnamefont {J.~E.}\ \bibnamefont
  {Moyal}},\ }\bibfield  {title} {\enquote {\bibinfo {title} {Quantum mechanics
  as a statistical theory},}\ }\href {\doibase 10.1017/S0305004100000487}
  {\bibfield  {journal} {\bibinfo  {journal} {Mathematical Proceedings of the
  Cambridge Philosophical Society}\ }\textbf {\bibinfo {volume} {45}},\
  \bibinfo {pages} {99–124} (\bibinfo {year} {1949})}\BibitemShut {NoStop}%
\bibitem [{\citenamefont {Stratonovich}(1957)}]{Stratonovich1957}%
  \BibitemOpen
  \bibfield  {author} {\bibinfo {author} {\bibfnamefont {R.~L.}\ \bibnamefont
  {Stratonovich}},\ }\bibfield  {title} {\enquote {\bibinfo {title} {On
  distributions in representation space},}\ }\href
  {http://jetp.ras.ru/cgi-bin/dn/e_004_06_0891.pdf} {\bibfield  {journal}
  {\bibinfo  {journal} {JETP}\ }\textbf {\bibinfo {volume} {4}},\ \bibinfo
  {pages} {891} (\bibinfo {year} {1957})}\BibitemShut {NoStop}%
\bibitem [{\citenamefont {Agarwal}\ and\ \citenamefont
  {Wolf}(1970{\natexlab{a}})}]{AgarwalPhysRevD.2.2161}%
  \BibitemOpen
  \bibfield  {author} {\bibinfo {author} {\bibfnamefont {G.~S.}\ \bibnamefont
  {Agarwal}}\ and\ \bibinfo {author} {\bibfnamefont {E.}~\bibnamefont {Wolf}},\
  }\bibfield  {title} {\enquote {\bibinfo {title} {Calculus for functions of
  noncommuting operators and general phase-space methods in quantum mechanics.
  i. mapping theorems and ordering of functions of noncommuting operators},}\
  }\href {\doibase 10.1103/PhysRevD.2.2161} {\bibfield  {journal} {\bibinfo
  {journal} {Phys. Rev. D}\ }\textbf {\bibinfo {volume} {2}},\ \bibinfo {pages}
  {2161--2186} (\bibinfo {year} {1970}{\natexlab{a}})}\BibitemShut {NoStop}%
\bibitem [{\citenamefont {Agarwal}\ and\ \citenamefont
  {Wolf}(1970{\natexlab{b}})}]{AgarwalPhysRevD.2.2187}%
  \BibitemOpen
  \bibfield  {author} {\bibinfo {author} {\bibfnamefont {G.~S.}\ \bibnamefont
  {Agarwal}}\ and\ \bibinfo {author} {\bibfnamefont {E.}~\bibnamefont {Wolf}},\
  }\bibfield  {title} {\enquote {\bibinfo {title} {Calculus for functions of
  noncommuting operators and general phase-space methods in quantum mechanics.
  ii. quantum mechanics in phase space},}\ }\href {\doibase
  10.1103/PhysRevD.2.2187} {\bibfield  {journal} {\bibinfo  {journal} {Phys.
  Rev. D}\ }\textbf {\bibinfo {volume} {2}},\ \bibinfo {pages} {2187--2205}
  (\bibinfo {year} {1970}{\natexlab{b}})}\BibitemShut {NoStop}%
\bibitem [{\citenamefont {Agarwal}\ and\ \citenamefont
  {Wolf}(1970{\natexlab{c}})}]{AgarwalPhysRevD.2.2206}%
  \BibitemOpen
  \bibfield  {author} {\bibinfo {author} {\bibfnamefont {G.~S.}\ \bibnamefont
  {Agarwal}}\ and\ \bibinfo {author} {\bibfnamefont {E.}~\bibnamefont {Wolf}},\
  }\bibfield  {title} {\enquote {\bibinfo {title} {Calculus for functions of
  noncommuting operators and general phase-space methods in quantum mechanics.
  iii. a generalized wick theorem and multitime mapping},}\ }\href {\doibase
  10.1103/PhysRevD.2.2206} {\bibfield  {journal} {\bibinfo  {journal} {Phys.
  Rev. D}\ }\textbf {\bibinfo {volume} {2}},\ \bibinfo {pages} {2206--2225}
  (\bibinfo {year} {1970}{\natexlab{c}})}\BibitemShut {NoStop}%
\bibitem [{\citenamefont {Cahill}\ and\ \citenamefont
  {Glauber}(1969{\natexlab{a}})}]{PhysRev.177.1857}%
  \BibitemOpen
  \bibfield  {author} {\bibinfo {author} {\bibfnamefont {K.~E.}\ \bibnamefont
  {Cahill}}\ and\ \bibinfo {author} {\bibfnamefont {R.~J.}\ \bibnamefont
  {Glauber}},\ }\bibfield  {title} {\enquote {\bibinfo {title} {Ordered
  expansions in boson amplitude operators},}\ }\href {\doibase
  10.1103/PhysRev.177.1857} {\bibfield  {journal} {\bibinfo  {journal} {Phys.
  Rev.}\ }\textbf {\bibinfo {volume} {177}},\ \bibinfo {pages} {1857--1881}
  (\bibinfo {year} {1969}{\natexlab{a}})}\BibitemShut {NoStop}%
\bibitem [{\citenamefont {Cahill}\ and\ \citenamefont
  {Glauber}(1969{\natexlab{b}})}]{PhysRev.177.1882}%
  \BibitemOpen
  \bibfield  {author} {\bibinfo {author} {\bibfnamefont {K.~E.}\ \bibnamefont
  {Cahill}}\ and\ \bibinfo {author} {\bibfnamefont {R.~J.}\ \bibnamefont
  {Glauber}},\ }\bibfield  {title} {\enquote {\bibinfo {title} {Density
  operators and quasiprobability distributions},}\ }\href {\doibase
  10.1103/PhysRev.177.1882} {\bibfield  {journal} {\bibinfo  {journal} {Phys.
  Rev.}\ }\textbf {\bibinfo {volume} {177}},\ \bibinfo {pages} {1882--1902}
  (\bibinfo {year} {1969}{\natexlab{b}})}\BibitemShut {NoStop}%
\bibitem [{\citenamefont {Berezin}(1975)}]{Berezin1975}%
  \BibitemOpen
  \bibfield  {author} {\bibinfo {author} {\bibfnamefont {F.~A.}\ \bibnamefont
  {Berezin}},\ }\bibfield  {title} {\enquote {\bibinfo {title} {General concept
  of quantization},}\ }\href {\doibase 10.1007/BF01609397} {\bibfield
  {journal} {\bibinfo  {journal} {Communications in Mathematical Physics}\
  }\textbf {\bibinfo {volume} {40}},\ \bibinfo {pages} {153--174} (\bibinfo
  {year} {1975})}\BibitemShut {NoStop}%
\bibitem [{\citenamefont {Várilly}\ and\ \citenamefont
  {Gracia-Bondía}(1989)}]{VARILLY1989107}%
  \BibitemOpen
  \bibfield  {author} {\bibinfo {author} {\bibfnamefont {Joseph~C}\
  \bibnamefont {Várilly}}\ and\ \bibinfo {author} {\bibfnamefont {JoséM}\
  \bibnamefont {Gracia-Bondía}},\ }\bibfield  {title} {\enquote {\bibinfo
  {title} {The moyal representation for spin},}\ }\href {\doibase
  10.1016/0003-4916(89)90262-5} {\bibfield  {journal} {\bibinfo  {journal}
  {Annals of Physics}\ }\textbf {\bibinfo {volume} {190}},\ \bibinfo {pages}
  {107–148} (\bibinfo {year} {1989})}\BibitemShut {NoStop}%
\bibitem [{\citenamefont {Klimov}\ and\ \citenamefont
  {Espinoza}(2002)}]{Klimov_2002}%
  \BibitemOpen
  \bibfield  {author} {\bibinfo {author} {\bibfnamefont {A~B}\ \bibnamefont
  {Klimov}}\ and\ \bibinfo {author} {\bibfnamefont {P}~\bibnamefont
  {Espinoza}},\ }\bibfield  {title} {\enquote {\bibinfo {title} {Moyal-like
  form of the star product for generalized su(2) stratonovich-weyl symbols},}\
  }\href {\doibase 10.1088/0305-4470/35/40/305} {\bibfield  {journal} {\bibinfo
   {journal} {Journal of Physics A: Mathematical and General}\ }\textbf
  {\bibinfo {volume} {35}},\ \bibinfo {pages} {8435} (\bibinfo {year}
  {2002})}\BibitemShut {NoStop}%
\bibitem [{\citenamefont {Agarwal}(1981)}]{AgarwalPhysRevA.24.2889}%
  \BibitemOpen
  \bibfield  {author} {\bibinfo {author} {\bibfnamefont {G.~S.}\ \bibnamefont
  {Agarwal}},\ }\bibfield  {title} {\enquote {\bibinfo {title} {Relation
  between atomic coherent-state representation, state multipoles, and
  generalized phase-space distributions},}\ }\href {\doibase
  10.1103/PhysRevA.24.2889} {\bibfield  {journal} {\bibinfo  {journal} {Phys.
  Rev. A}\ }\textbf {\bibinfo {volume} {24}},\ \bibinfo {pages} {2889–2896}
  (\bibinfo {year} {1981})}\BibitemShut {NoStop}%
\bibitem [{\citenamefont {Radcliffe}(1971)}]{JMRadcliffe_1971}%
  \BibitemOpen
  \bibfield  {author} {\bibinfo {author} {\bibfnamefont {J~M}\ \bibnamefont
  {Radcliffe}},\ }\bibfield  {title} {\enquote {\bibinfo {title} {Some
  properties of coherent spin states},}\ }\href {\doibase
  10.1088/0305-4470/4/3/009} {\bibfield  {journal} {\bibinfo  {journal}
  {Journal of Physics A: General Physics}\ }\textbf {\bibinfo {volume} {4}},\
  \bibinfo {pages} {313} (\bibinfo {year} {1971})}\BibitemShut {NoStop}%
\bibitem [{\citenamefont {Arecchi}\ \emph {et~al.}(1972)\citenamefont
  {Arecchi}, \citenamefont {Courtens}, \citenamefont {Gilmore},\ and\
  \citenamefont {Thomas}}]{GilmorePhysRevA.6.2211}%
  \BibitemOpen
  \bibfield  {author} {\bibinfo {author} {\bibfnamefont {F.~T.}\ \bibnamefont
  {Arecchi}}, \bibinfo {author} {\bibfnamefont {Eric}\ \bibnamefont
  {Courtens}}, \bibinfo {author} {\bibfnamefont {Robert}\ \bibnamefont
  {Gilmore}}, \ and\ \bibinfo {author} {\bibfnamefont {Harry}\ \bibnamefont
  {Thomas}},\ }\bibfield  {title} {\enquote {\bibinfo {title} {Atomic coherent
  states in quantum optics},}\ }\href {\doibase 10.1103/PhysRevA.6.2211}
  {\bibfield  {journal} {\bibinfo  {journal} {Phys. Rev. A}\ }\textbf {\bibinfo
  {volume} {6}},\ \bibinfo {pages} {2211–2237} (\bibinfo {year}
  {1972})}\BibitemShut {NoStop}%
\bibitem [{\citenamefont {Gilmore}(1979)}]{GilmoreJMP}%
  \BibitemOpen
  \bibfield  {author} {\bibinfo {author} {\bibfnamefont {R.}~\bibnamefont
  {Gilmore}},\ }\bibfield  {title} {\enquote {\bibinfo {title} {The classical
  limit of quantum nonspin systems},}\ }\href {\doibase 10.1063/1.524137}
  {\bibfield  {journal} {\bibinfo  {journal} {Journal of Mathematical Physics}\
  }\textbf {\bibinfo {volume} {20}},\ \bibinfo {pages} {891–893} (\bibinfo
  {year} {1979})},\ \Eprint
  {http://arxiv.org/abs/https://doi.org/10.1063/1.524137}
  {https://doi.org/10.1063/1.524137} \BibitemShut {NoStop}%
\bibitem [{\citenamefont {Schr{\"o}dinger}(1926)}]{Schrodinger1926}%
  \BibitemOpen
  \bibfield  {author} {\bibinfo {author} {\bibfnamefont {E.}~\bibnamefont
  {Schr{\"o}dinger}},\ }\bibfield  {title} {\enquote {\bibinfo {title} {Der
  stetige {\"u}bergang von der mikro- zur makromechanik},}\ }\href {\doibase
  10.1007/BF01507634} {\bibfield  {journal} {\bibinfo  {journal}
  {Naturwissenschaften}\ }\textbf {\bibinfo {volume} {14}},\ \bibinfo {pages}
  {664--666} (\bibinfo {year} {1926})}\BibitemShut {NoStop}%
\bibitem [{\citenamefont {Klauder}(1960)}]{KLAUDER1960123}%
  \BibitemOpen
  \bibfield  {author} {\bibinfo {author} {\bibfnamefont {John~R}\ \bibnamefont
  {Klauder}},\ }\bibfield  {title} {\enquote {\bibinfo {title} {The action
  option and a feynman quantization of spinor fields in terms of ordinary
  c-numbers},}\ }\href {\doibase https://doi.org/10.1016/0003-4916(60)90131-7}
  {\bibfield  {journal} {\bibinfo  {journal} {Annals of Physics}\ }\textbf
  {\bibinfo {volume} {11}},\ \bibinfo {pages} {123--168} (\bibinfo {year}
  {1960})}\BibitemShut {NoStop}%
\bibitem [{\citenamefont {Gilmore}(1981)}]{Gilmorebook}%
  \BibitemOpen
  \bibfield  {author} {\bibinfo {author} {\bibfnamefont {R.}~\bibnamefont
  {Gilmore}},\ }\href
  {https://onlinelibrary.wiley.com/doi/abs/10.1002/3527600434.eap052.pub2}
  {\emph {\bibinfo {title} {Catastrophe Theory for Scientists and Engineers}}}\
  (\bibinfo  {publisher} {Wiley, New York},\ \bibinfo {year}
  {1981})\BibitemShut {NoStop}%
\bibitem [{\citenamefont {Klauder}\ and\ \citenamefont
  {Skagerstam}(1985)}]{Klauderbook}%
  \BibitemOpen
  \bibfield  {author} {\bibinfo {author} {\bibfnamefont {J}~\bibnamefont
  {Klauder}}\ and\ \bibinfo {author} {\bibfnamefont {B}~\bibnamefont
  {Skagerstam}},\ }\href {\doibase 10.1142/0096} {\emph {\bibinfo {title}
  {Coherent States}}}\ (\bibinfo  {publisher} {World Scientific},\ \bibinfo
  {year} {1985})\ \Eprint
  {http://arxiv.org/abs/https://www.worldscientific.com/doi/pdf/10.1142/0096}
  {https://www.worldscientific.com/doi/pdf/10.1142/0096} \BibitemShut {NoStop}%
\bibitem [{\citenamefont {Perelomov}(1986)}]{Perelomov}%
  \BibitemOpen
  \bibfield  {author} {\bibinfo {author} {\bibfnamefont {Askold}\ \bibnamefont
  {Perelomov}},\ }\href {\doibase 10.1007/978-3-642-61629-7} {\emph {\bibinfo
  {title} {Generalized Coherent States and their Applications}}}\ (\bibinfo
  {publisher} {Springer-Verlag Berlin Heidelberg},\ \bibinfo {year}
  {1986})\BibitemShut {NoStop}%
\bibitem [{\citenamefont {Zhang}\ \emph {et~al.}(1990)\citenamefont {Zhang},
  \citenamefont {Feng},\ and\ \citenamefont {Gilmore}}]{RevModPhys.62.867}%
  \BibitemOpen
  \bibfield  {author} {\bibinfo {author} {\bibfnamefont {Wei-Min}\ \bibnamefont
  {Zhang}}, \bibinfo {author} {\bibfnamefont {Da~Hsuan}\ \bibnamefont {Feng}},
  \ and\ \bibinfo {author} {\bibfnamefont {Robert}\ \bibnamefont {Gilmore}},\
  }\bibfield  {title} {\enquote {\bibinfo {title} {Coherent states: Theory and
  some applications},}\ }\href {\doibase 10.1103/RevModPhys.62.867} {\bibfield
  {journal} {\bibinfo  {journal} {Rev. Mod. Phys.}\ }\textbf {\bibinfo {volume}
  {62}},\ \bibinfo {pages} {867–927} (\bibinfo {year} {1990})}\BibitemShut
  {NoStop}%
\bibitem [{\citenamefont {Gazeau}(2009)}]{Gazeaubook}%
  \BibitemOpen
  \bibfield  {author} {\bibinfo {author} {\bibfnamefont {J.P.}\ \bibnamefont
  {Gazeau}},\ }\href {\doibase 10.1002/9783527628285} {\emph {\bibinfo {title}
  {Coherent States in Quantum Physics}}}\ (\bibinfo  {publisher} {John Wiley \&
  Sons, Ltd},\ \bibinfo {year} {2009})\BibitemShut {NoStop}%
\bibitem [{\citenamefont {Brif}\ and\ \citenamefont {Mann}(1998)}]{Brif_1998}%
  \BibitemOpen
  \bibfield  {author} {\bibinfo {author} {\bibfnamefont {C}~\bibnamefont
  {Brif}}\ and\ \bibinfo {author} {\bibfnamefont {A}~\bibnamefont {Mann}},\
  }\bibfield  {title} {\enquote {\bibinfo {title} {A general theory of
  phase-space quasiprobability distributions},}\ }\href {\doibase
  10.1088/0305-4470/31/1/002} {\bibfield  {journal} {\bibinfo  {journal}
  {Journal of Physics A: Mathematical and General}\ }\textbf {\bibinfo {volume}
  {31}},\ \bibinfo {pages} {L9–L17} (\bibinfo {year} {1998})}\BibitemShut
  {NoStop}%
\bibitem [{\citenamefont {Brif}\ and\ \citenamefont
  {Mann}(1999)}]{PhysRevA.59.971}%
  \BibitemOpen
  \bibfield  {author} {\bibinfo {author} {\bibfnamefont {C.}~\bibnamefont
  {Brif}}\ and\ \bibinfo {author} {\bibfnamefont {A.}~\bibnamefont {Mann}},\
  }\bibfield  {title} {\enquote {\bibinfo {title} {Phase-space formulation of
  quantum mechanics and quantum-state reconstruction for physical systems with
  {L}ie-group symmetries},}\ }\href {\doibase 10.1103/PhysRevA.59.971}
  {\bibfield  {journal} {\bibinfo  {journal} {Phys. Rev. A}\ }\textbf {\bibinfo
  {volume} {59}},\ \bibinfo {pages} {971–987} (\bibinfo {year}
  {1999})}\BibitemShut {NoStop}%
\bibitem [{\citenamefont {Ali}\ \emph {et~al.}(2000)\citenamefont {Ali},
  \citenamefont {Atakishiyev}, \citenamefont {Chumakov},\ and\ \citenamefont
  {Wolf}}]{Ali2000}%
  \BibitemOpen
  \bibfield  {author} {\bibinfo {author} {\bibfnamefont {S.~T.}\ \bibnamefont
  {Ali}}, \bibinfo {author} {\bibfnamefont {N.~M.}\ \bibnamefont
  {Atakishiyev}}, \bibinfo {author} {\bibfnamefont {S.~M.}\ \bibnamefont
  {Chumakov}}, \ and\ \bibinfo {author} {\bibfnamefont {K.~B.}\ \bibnamefont
  {Wolf}},\ }\bibfield  {title} {\enquote {\bibinfo {title} {The wigner
  function for general lie groups and the wavelet transform},}\ }\href
  {\doibase 10.1007/PL00001012} {\bibfield  {journal} {\bibinfo  {journal}
  {Annales Henri Poincar{\'e}}\ }\textbf {\bibinfo {volume} {1}},\ \bibinfo
  {pages} {685--714} (\bibinfo {year} {2000})}\BibitemShut {NoStop}%
\bibitem [{\citenamefont {Martins}\ \emph {et~al.}(2019)\citenamefont
  {Martins}, \citenamefont {Klimov},\ and\ \citenamefont
  {de~Guise}}]{MartinsKlimovGuise_SU3}%
  \BibitemOpen
  \bibfield  {author} {\bibinfo {author} {\bibfnamefont {Alex Clésio~Nunes}\
  \bibnamefont {Martins}}, \bibinfo {author} {\bibfnamefont {Andrei~B}\
  \bibnamefont {Klimov}}, \ and\ \bibinfo {author} {\bibfnamefont {Hubert}\
  \bibnamefont {de~Guise}},\ }\bibfield  {title} {\enquote {\bibinfo {title}
  {Correspondence rules for wigner functions over su(3)/u(2)},}\ }\href
  {\doibase 10.1088/1751-8121/ab226c} {\bibfield  {journal} {\bibinfo
  {journal} {Journal of Physics A: Mathematical and Theoretical}\ }\textbf
  {\bibinfo {volume} {52}},\ \bibinfo {pages} {285202} (\bibinfo {year}
  {2019})}\BibitemShut {NoStop}%
\bibitem [{\citenamefont {Klimov}\ and\ \citenamefont
  {de~Guise}(2010)}]{Klimov_2010}%
  \BibitemOpen
  \bibfield  {author} {\bibinfo {author} {\bibfnamefont {Andrei~B}\
  \bibnamefont {Klimov}}\ and\ \bibinfo {author} {\bibfnamefont {Hubert}\
  \bibnamefont {de~Guise}},\ }\bibfield  {title} {\enquote {\bibinfo {title}
  {General approach to {SU(n)} quasi-distribution functions},}\ }\href
  {\doibase 10.1088/1751-8113/43/40/402001} {\bibfield  {journal} {\bibinfo
  {journal} {Journal of Physics A: Mathematical and Theoretical}\ }\textbf
  {\bibinfo {volume} {43}},\ \bibinfo {pages} {402001} (\bibinfo {year}
  {2010})}\BibitemShut {NoStop}%
\bibitem [{\citenamefont {Klimov}\ \emph {et~al.}(2021)\citenamefont {Klimov},
  \citenamefont {Seyfarth}, \citenamefont {de~Guise},\ and\ \citenamefont
  {Sánchez-Soto}}]{Klimov_2021}%
  \BibitemOpen
  \bibfield  {author} {\bibinfo {author} {\bibfnamefont {Andrei~B}\
  \bibnamefont {Klimov}}, \bibinfo {author} {\bibfnamefont {Ulrich}\
  \bibnamefont {Seyfarth}}, \bibinfo {author} {\bibfnamefont {Hubert}\
  \bibnamefont {de~Guise}}, \ and\ \bibinfo {author} {\bibfnamefont {Luis~L}\
  \bibnamefont {Sánchez-Soto}},\ }\bibfield  {title} {\enquote {\bibinfo
  {title} {Su(1, 1) covariant s-parametrized maps},}\ }\href {\doibase
  10.1088/1751-8121/abd7b4} {\bibfield  {journal} {\bibinfo  {journal} {Journal
  of Physics A: Mathematical and Theoretical}\ }\textbf {\bibinfo {volume}
  {54}},\ \bibinfo {pages} {065301} (\bibinfo {year} {2021})}\BibitemShut
  {NoStop}%
\bibitem [{\citenamefont {Atakishiyev}\ \emph {et~al.}(1998)\citenamefont
  {Atakishiyev}, \citenamefont {Chumakov},\ and\ \citenamefont
  {Wolf}}]{Atakishiyev1998-2}%
  \BibitemOpen
  \bibfield  {author} {\bibinfo {author} {\bibfnamefont {Natig~M.}\
  \bibnamefont {Atakishiyev}}, \bibinfo {author} {\bibfnamefont {Sergey~M.}\
  \bibnamefont {Chumakov}}, \ and\ \bibinfo {author} {\bibfnamefont
  {Kurt~Bernardo}\ \bibnamefont {Wolf}},\ }\bibfield  {title} {\enquote
  {\bibinfo {title} {Wigner distribution function for finite systems},}\ }\href
  {\doibase 10.1063/1.532636} {\bibfield  {journal} {\bibinfo  {journal}
  {Journal of Mathematical Physics}\ }\textbf {\bibinfo {volume} {39}},\
  \bibinfo {pages} {6247--6261} (\bibinfo {year} {1998})},\ \Eprint
  {http://arxiv.org/abs/https://pubs.aip.org/aip/jmp/article-pdf/39/12/6247/19291270/6247\_1\_online.pdf}
  {https://pubs.aip.org/aip/jmp/article-pdf/39/12/6247/19291270/6247\_1\_online.pdf}
  \BibitemShut {NoStop}%
\bibitem [{\citenamefont {Tilma}\ \emph {et~al.}(2016)\citenamefont {Tilma},
  \citenamefont {Everitt}, \citenamefont {Samson}, \citenamefont {Munro},\ and\
  \citenamefont {Nemoto}}]{TilmaPhysRevLett.117.180401}%
  \BibitemOpen
  \bibfield  {author} {\bibinfo {author} {\bibfnamefont {Todd}\ \bibnamefont
  {Tilma}}, \bibinfo {author} {\bibfnamefont {Mark~J.}\ \bibnamefont
  {Everitt}}, \bibinfo {author} {\bibfnamefont {John~H.}\ \bibnamefont
  {Samson}}, \bibinfo {author} {\bibfnamefont {William~J.}\ \bibnamefont
  {Munro}}, \ and\ \bibinfo {author} {\bibfnamefont {Kae}\ \bibnamefont
  {Nemoto}},\ }\bibfield  {title} {\enquote {\bibinfo {title} {Wigner functions
  for arbitrary quantum systems},}\ }\href {\doibase
  10.1103/PhysRevLett.117.180401} {\bibfield  {journal} {\bibinfo  {journal}
  {Phys. Rev. Lett.}\ }\textbf {\bibinfo {volume} {117}},\ \bibinfo {pages}
  {180401} (\bibinfo {year} {2016})}\BibitemShut {NoStop}%
\bibitem [{\citenamefont {Dodonov}\ \emph {et~al.}(1974)\citenamefont
  {Dodonov}, \citenamefont {Malkin},\ and\ \citenamefont
  {Man'ko}}]{Dodonovcat}%
  \BibitemOpen
  \bibfield  {author} {\bibinfo {author} {\bibfnamefont {V.V.}\ \bibnamefont
  {Dodonov}}, \bibinfo {author} {\bibfnamefont {I.A.}\ \bibnamefont {Malkin}},
  \ and\ \bibinfo {author} {\bibfnamefont {V.I.}\ \bibnamefont {Man'ko}},\
  }\bibfield  {title} {\enquote {\bibinfo {title} {Even and odd coherent states
  and excitations of a singular oscillator},}\ }\href {\doibase
  10.1016/0031-8914(74)90215-8} {\bibfield  {journal} {\bibinfo  {journal}
  {Physica}\ }\textbf {\bibinfo {volume} {72}},\ \bibinfo {pages} {597–615}
  (\bibinfo {year} {1974})}\BibitemShut {NoStop}%
\bibitem [{\citenamefont {Nieto}\ and\ \citenamefont
  {Truax}(1993)}]{M.Nieto&D.Traux}%
  \BibitemOpen
  \bibfield  {author} {\bibinfo {author} {\bibfnamefont {Michael~Martin}\
  \bibnamefont {Nieto}}\ and\ \bibinfo {author} {\bibfnamefont {D.~Rodney}\
  \bibnamefont {Truax}},\ }\bibfield  {title} {\enquote {\bibinfo {title}
  {Squeezed states for general systems},}\ }\href {\doibase
  10.1103/PhysRevLett.71.2843} {\bibfield  {journal} {\bibinfo  {journal}
  {Phys. Rev. Lett.}\ }\textbf {\bibinfo {volume} {71}},\ \bibinfo {pages}
  {2843–2846} (\bibinfo {year} {1993})}\BibitemShut {NoStop}%
\bibitem [{\citenamefont {{Bu\ifmmode \check{z}\else ž\fi{}ek}}\ \emph
  {et~al.}(1992)\citenamefont {{Bu\ifmmode \check{z}\else ž\fi{}ek}},
  \citenamefont {Vidiella-Barranco},\ and\ \citenamefont
  {Knight}}]{V.Buzek&A.Viiella-Barranco&P.Knight}%
  \BibitemOpen
  \bibfield  {author} {\bibinfo {author} {\bibfnamefont {V.}~\bibnamefont
  {{Bu\ifmmode \check{z}\else ž\fi{}ek}}}, \bibinfo {author} {\bibfnamefont
  {A.}~\bibnamefont {Vidiella-Barranco}}, \ and\ \bibinfo {author}
  {\bibfnamefont {P.~L.}\ \bibnamefont {Knight}},\ }\bibfield  {title}
  {\enquote {\bibinfo {title} {Superpositions of coherent states: Squeezing and
  dissipation},}\ }\href {\doibase 10.1103/PhysRevA.45.6570} {\bibfield
  {journal} {\bibinfo  {journal} {Phys. Rev. A}\ }\textbf {\bibinfo {volume}
  {45}},\ \bibinfo {pages} {6570–6585} (\bibinfo {year} {1992})}\BibitemShut
  {NoStop}%
\bibitem [{\citenamefont {Hillery}(1987)}]{M.Hillery}%
  \BibitemOpen
  \bibfield  {author} {\bibinfo {author} {\bibfnamefont {Mark}\ \bibnamefont
  {Hillery}},\ }\bibfield  {title} {\enquote {\bibinfo {title}
  {Amplitude-squared squeezing of the electromagnetic field},}\ }\href
  {\doibase 10.1103/PhysRevA.36.3796} {\bibfield  {journal} {\bibinfo
  {journal} {Phys. Rev. A}\ }\textbf {\bibinfo {volume} {36}},\ \bibinfo
  {pages} {3796–3802} (\bibinfo {year} {1987})}\BibitemShut {NoStop}%
\bibitem [{\citenamefont {Casta\~nos}\ \emph {et~al.}(1995)\citenamefont
  {Casta\~nos}, \citenamefont {López-Pe\~na},\ and\ \citenamefont
  {Man'ko}}]{Mankocat}%
  \BibitemOpen
  \bibfield  {author} {\bibinfo {author} {\bibfnamefont {O.}~\bibnamefont
  {Casta\~nos}}, \bibinfo {author} {\bibfnamefont {R.}~\bibnamefont
  {López-Pe\~na}}, \ and\ \bibinfo {author} {\bibfnamefont {V.~I.}\
  \bibnamefont {Man'ko}},\ }\bibfield  {title} {\enquote {\bibinfo {title}
  {Crystallized schrödinger cat states},}\ }\href {\doibase
  10.1007/BF02581033} {\bibfield  {journal} {\bibinfo  {journal} {Journal of
  Russian Laser Research}\ }\textbf {\bibinfo {volume} {16}},\ \bibinfo {pages}
  {477–525} (\bibinfo {year} {1995})}\BibitemShut {NoStop}%
\bibitem [{\citenamefont {Pezzè}\ \emph {et~al.}(2018)\citenamefont {Pezzè},
  \citenamefont {Smerzi}, \citenamefont {Oberthaler}, \citenamefont {Schmied},\
  and\ \citenamefont {Treutlein}}]{RevModPhys.90.035005-Pezze}%
  \BibitemOpen
  \bibfield  {author} {\bibinfo {author} {\bibfnamefont {Luca}\ \bibnamefont
  {Pezzè}}, \bibinfo {author} {\bibfnamefont {Augusto}\ \bibnamefont
  {Smerzi}}, \bibinfo {author} {\bibfnamefont {Markus~K.}\ \bibnamefont
  {Oberthaler}}, \bibinfo {author} {\bibfnamefont {Roman}\ \bibnamefont
  {Schmied}}, \ and\ \bibinfo {author} {\bibfnamefont {Philipp}\ \bibnamefont
  {Treutlein}},\ }\bibfield  {title} {\enquote {\bibinfo {title} {Quantum
  metrology with nonclassical states of atomic ensembles},}\ }\href {\doibase
  10.1103/RevModPhys.90.035005} {\bibfield  {journal} {\bibinfo  {journal}
  {Rev. Mod. Phys.}\ }\textbf {\bibinfo {volume} {90}},\ \bibinfo {pages}
  {035005} (\bibinfo {year} {2018})}\BibitemShut {NoStop}%
\bibitem [{\citenamefont {Agarwal}\ \emph {et~al.}(1997)\citenamefont
  {Agarwal}, \citenamefont {Puri},\ and\ \citenamefont
  {Singh}}]{Agarwal_PhysRevA.56.2249}%
  \BibitemOpen
  \bibfield  {author} {\bibinfo {author} {\bibfnamefont {G.~S.}\ \bibnamefont
  {Agarwal}}, \bibinfo {author} {\bibfnamefont {R.~R.}\ \bibnamefont {Puri}}, \
  and\ \bibinfo {author} {\bibfnamefont {R.~P.}\ \bibnamefont {Singh}},\
  }\bibfield  {title} {\enquote {\bibinfo {title} {Atomic {Schr\"odinger} cat
  states},}\ }\href {\doibase 10.1103/PhysRevA.56.2249} {\bibfield  {journal}
  {\bibinfo  {journal} {Phys. Rev. A}\ }\textbf {\bibinfo {volume} {56}},\
  \bibinfo {pages} {2249--2254} (\bibinfo {year} {1997})}\BibitemShut {NoStop}%
\bibitem [{\citenamefont {Kitagawa}\ and\ \citenamefont
  {Ueda}(1993)}]{Kitagawa}%
  \BibitemOpen
  \bibfield  {author} {\bibinfo {author} {\bibfnamefont {Masahiro}\
  \bibnamefont {Kitagawa}}\ and\ \bibinfo {author} {\bibfnamefont {Masahito}\
  \bibnamefont {Ueda}},\ }\bibfield  {title} {\enquote {\bibinfo {title}
  {Squeezed spin states},}\ }\href {\doibase 10.1103/PhysRevA.47.5138}
  {\bibfield  {journal} {\bibinfo  {journal} {Phys. Rev. A}\ }\textbf {\bibinfo
  {volume} {47}},\ \bibinfo {pages} {5138–5143} (\bibinfo {year}
  {1993})}\BibitemShut {NoStop}%
\bibitem [{\citenamefont {Calixto}\ \emph {et~al.}(2017)\citenamefont
  {Calixto}, \citenamefont {Castaños},\ and\ \citenamefont
  {Romera}}]{Calixto_2017}%
  \BibitemOpen
  \bibfield  {author} {\bibinfo {author} {\bibfnamefont {Manuel}\ \bibnamefont
  {Calixto}}, \bibinfo {author} {\bibfnamefont {Octavio}\ \bibnamefont
  {Castaños}}, \ and\ \bibinfo {author} {\bibfnamefont {Elvira}\ \bibnamefont
  {Romera}},\ }\bibfield  {title} {\enquote {\bibinfo {title} {Entanglement and
  quantum phase diagrams of symmetric multi-qubit systems},}\ }\href {\doibase
  10.1088/1742-5468/aa8703} {\bibfield  {journal} {\bibinfo  {journal} {Journal
  of Statistical Mechanics: Theory and Experiment}\ }\textbf {\bibinfo {volume}
  {2017}},\ \bibinfo {pages} {103103} (\bibinfo {year} {2017})}\BibitemShut
  {NoStop}%
\bibitem [{\citenamefont {Calixto}\ \emph
  {et~al.}(2021{\natexlab{a}})\citenamefont {Calixto}, \citenamefont
  {Mayorgas},\ and\ \citenamefont {Guerrero}}]{nuestroPRE}%
  \BibitemOpen
  \bibfield  {author} {\bibinfo {author} {\bibfnamefont {Manuel}\ \bibnamefont
  {Calixto}}, \bibinfo {author} {\bibfnamefont {Alberto}\ \bibnamefont
  {Mayorgas}}, \ and\ \bibinfo {author} {\bibfnamefont {Julio}\ \bibnamefont
  {Guerrero}},\ }\bibfield  {title} {\enquote {\bibinfo {title} {Role of mixed
  permutation symmetry sectors in the thermodynamic limit of critical
  three-level {Lipkin-Meshkov-Glick} atom models},}\ }\href {\doibase
  10.1103/PhysRevE.103.012116} {\bibfield  {journal} {\bibinfo  {journal}
  {Phys. Rev. E}\ }\textbf {\bibinfo {volume} {103}},\ \bibinfo {pages}
  {012116} (\bibinfo {year} {2021}{\natexlab{a}})}\BibitemShut {NoStop}%
\bibitem [{\citenamefont {Calixto}\ \emph
  {et~al.}(2021{\natexlab{b}})\citenamefont {Calixto}, \citenamefont
  {Mayorgas},\ and\ \citenamefont {Guerrero}}]{QIP-2021-Entanglement}%
  \BibitemOpen
  \bibfield  {author} {\bibinfo {author} {\bibfnamefont {Manuel}\ \bibnamefont
  {Calixto}}, \bibinfo {author} {\bibfnamefont {Alberto}\ \bibnamefont
  {Mayorgas}}, \ and\ \bibinfo {author} {\bibfnamefont {Julio}\ \bibnamefont
  {Guerrero}},\ }\bibfield  {title} {\enquote {\bibinfo {title} {Entanglement
  and {U(D)}-spin squeezing in symmetric multi-qudit systems and applications
  to quantum phase transitions in {L}ipkin–{M}eshkov–{G}lick {D}-level atom
  models},}\ }\href {\doibase 10.1007/s11128-021-03218-6} {\bibfield  {journal}
  {\bibinfo  {journal} {Quantum Information Processing}\ }\textbf {\bibinfo
  {volume} {20}},\ \bibinfo {pages} {304} (\bibinfo {year}
  {2021}{\natexlab{b}})}\BibitemShut {NoStop}%
\bibitem [{\citenamefont {Guerrero}\ \emph {et~al.}(2022)\citenamefont
  {Guerrero}, \citenamefont {Mayorgas},\ and\ \citenamefont
  {Calixto}}]{Guerrero22}%
  \BibitemOpen
  \bibfield  {author} {\bibinfo {author} {\bibfnamefont {J.}~\bibnamefont
  {Guerrero}}, \bibinfo {author} {\bibfnamefont {A.}~\bibnamefont {Mayorgas}},
  \ and\ \bibinfo {author} {\bibfnamefont {M.}~\bibnamefont {Calixto}},\
  }\bibfield  {title} {\enquote {\bibinfo {title} {Information diagrams in the
  study of entanglement in symmetric multi-qudit systems and applications to
  quantum phase transitions in {Lipkin}–{Meshkov}–{Glick} {D}-level atom
  models},}\ }\href {\doibase 10.1007/s11128-022-03524-7} {\bibfield  {journal}
  {\bibinfo  {journal} {Quant. Inf. Process.}\ }\textbf {\bibinfo {volume}
  {21}},\ \bibinfo {pages} {223} (\bibinfo {year} {2022})}\BibitemShut
  {NoStop}%
\bibitem [{\citenamefont {Guerrero}\ \emph {et~al.}(2023)\citenamefont
  {Guerrero}, \citenamefont {Sojo}, \citenamefont {Mayorgas},\ and\
  \citenamefont {Calixto}}]{Guerrero_2023}%
  \BibitemOpen
  \bibfield  {author} {\bibinfo {author} {\bibfnamefont {Julio}\ \bibnamefont
  {Guerrero}}, \bibinfo {author} {\bibfnamefont {Antonio}\ \bibnamefont
  {Sojo}}, \bibinfo {author} {\bibfnamefont {Alberto}\ \bibnamefont
  {Mayorgas}}, \ and\ \bibinfo {author} {\bibfnamefont {Manuel}\ \bibnamefont
  {Calixto}},\ }\bibfield  {title} {\enquote {\bibinfo {title} {Schmidt
  decomposition of parity adapted coherent states for symmetric
  multi-qudits},}\ }\href {\doibase 10.1088/1751-8121/aceae0} {\bibfield
  {journal} {\bibinfo  {journal} {Journal of Physics A: Mathematical and
  Theoretical}\ }\textbf {\bibinfo {volume} {56}},\ \bibinfo {pages} {355304}
  (\bibinfo {year} {2023})}\BibitemShut {NoStop}%
\bibitem [{\citenamefont {Mayorgas}\ \emph {et~al.}(2023)\citenamefont
  {Mayorgas}, \citenamefont {Guerrero},\ and\ \citenamefont
  {Calixto}}]{Mayorgas2023}%
  \BibitemOpen
  \bibfield  {author} {\bibinfo {author} {\bibfnamefont {Alberto}\ \bibnamefont
  {Mayorgas}}, \bibinfo {author} {\bibfnamefont {Julio}\ \bibnamefont
  {Guerrero}}, \ and\ \bibinfo {author} {\bibfnamefont {Manuel}\ \bibnamefont
  {Calixto}},\ }\bibfield  {title} {\enquote {\bibinfo {title} {{Localization
  measures of parity adapted {U($D$)}-spin coherent states applied to the phase
  space analysis of the {$D$}-level Lipkin-Meshkov-Glick model}},}\ }\href
  {\doibase 10.48550/arxiv.2302.06254} {\  (\bibinfo {year} {2023}),\
  10.48550/arxiv.2302.06254}\BibitemShut {NoStop}%
\bibitem [{\citenamefont {Qikeng}(1998)}]{Qikeng}%
  \BibitemOpen
  \bibfield  {author} {\bibinfo {author} {\bibfnamefont {Lu}~\bibnamefont
  {Qikeng}},\ }\bibfield  {title} {\enquote {\bibinfo {title} {The eigen
  functions of the complex projective space},}\ }\href {\doibase
  10.1007/BF02563877} {\bibfield  {journal} {\bibinfo  {journal} {Acta
  Mathematica Sinica}\ }\textbf {\bibinfo {volume} {14}},\ \bibinfo {eid} {1}
  (\bibinfo {year} {1998})}\BibitemShut {NoStop}%
\bibitem [{\citenamefont {Berger}\ \emph {et~al.}(1971)\citenamefont {Berger},
  \citenamefont {Gauduchon},\ and\ \citenamefont {Mazet}}]{BergerBook}%
  \BibitemOpen
  \bibfield  {author} {\bibinfo {author} {\bibfnamefont {Marcel}\ \bibnamefont
  {Berger}}, \bibinfo {author} {\bibfnamefont {Paul}\ \bibnamefont
  {Gauduchon}}, \ and\ \bibinfo {author} {\bibfnamefont {Edmond}\ \bibnamefont
  {Mazet}},\ }\href@noop {} {\emph {\bibinfo {title} {Le Spectre de une Variete
  Riemannienne}}},\ Lecture Notes in Mathematics, 194\ (\bibinfo  {publisher}
  {Springer Berlin Heidelberg},\ \bibinfo {address} {Berlin, Heidelberg},\
  \bibinfo {year} {1971})\BibitemShut {NoStop}%
\bibitem [{\citenamefont {Kenfack}\ and\ \citenamefont
  {Życzkowski}(2004)}]{AnatoleKenfack_2004}%
  \BibitemOpen
  \bibfield  {author} {\bibinfo {author} {\bibfnamefont {Anatole}\ \bibnamefont
  {Kenfack}}\ and\ \bibinfo {author} {\bibfnamefont {Karol}\ \bibnamefont
  {Życzkowski}},\ }\bibfield  {title} {\enquote {\bibinfo {title} {Negativity
  of the wigner function as an indicator of non-classicality},}\ }\href
  {\doibase 10.1088/1464-4266/6/10/003} {\bibfield  {journal} {\bibinfo
  {journal} {Journal of Optics B: Quantum and Semiclassical Optics}\ }\textbf
  {\bibinfo {volume} {6}},\ \bibinfo {pages} {396} (\bibinfo {year}
  {2004})}\BibitemShut {NoStop}%
\bibitem [{\citenamefont {Davis}\ \emph {et~al.}(2021)\citenamefont {Davis},
  \citenamefont {Kumari}, \citenamefont {Mann},\ and\ \citenamefont
  {Ghose}}]{PhysRevResearch.3.033134}%
  \BibitemOpen
  \bibfield  {author} {\bibinfo {author} {\bibfnamefont {Jack}\ \bibnamefont
  {Davis}}, \bibinfo {author} {\bibfnamefont {Meenu}\ \bibnamefont {Kumari}},
  \bibinfo {author} {\bibfnamefont {Robert~B.}\ \bibnamefont {Mann}}, \ and\
  \bibinfo {author} {\bibfnamefont {Shohini}\ \bibnamefont {Ghose}},\
  }\bibfield  {title} {\enquote {\bibinfo {title} {Wigner negativity in
  spin-$j$ systems},}\ }\href {\doibase 10.1103/PhysRevResearch.3.033134}
  {\bibfield  {journal} {\bibinfo  {journal} {Phys. Rev. Res.}\ }\textbf
  {\bibinfo {volume} {3}},\ \bibinfo {pages} {033134} (\bibinfo {year}
  {2021})}\BibitemShut {NoStop}%
\bibitem [{\citenamefont {Ansari}\ and\ \citenamefont
  {Man'ko}(1994)}]{Manko-MultimodeCats}%
  \BibitemOpen
  \bibfield  {author} {\bibinfo {author} {\bibfnamefont {N.~A.}\ \bibnamefont
  {Ansari}}\ and\ \bibinfo {author} {\bibfnamefont {V.I.}\ \bibnamefont
  {Man'ko}},\ }\bibfield  {title} {\enquote {\bibinfo {title} {Photon
  statistics of multimode even and odd coherent light},}\ }\href@noop {}
  {\bibfield  {journal} {\bibinfo  {journal} {Phys. Rev. A}\ }\textbf {\bibinfo
  {volume} {50}},\ \bibinfo {pages} {1942} (\bibinfo {year}
  {1994})}\BibitemShut {NoStop}%
\bibitem [{\citenamefont {Barut}\ and\ \citenamefont {Raczka}(1980)}]{Barut}%
  \BibitemOpen
  \bibfield  {author} {\bibinfo {author} {\bibfnamefont {A.O.}\ \bibnamefont
  {Barut}}\ and\ \bibinfo {author} {\bibfnamefont {R.}~\bibnamefont {Raczka}},\
  }\href {https://www.worldscientific.com/doi/10.1142/0352} {\emph {\bibinfo
  {title} {Theory of Group Representations and Applications}}}\ (\bibinfo
  {publisher} {Polish Scientific Publishers, Warszawa},\ \bibinfo {year}
  {1980})\BibitemShut {NoStop}%
\bibitem [{\citenamefont {Vourdas}(2006)}]{Vourdas_2006}%
  \BibitemOpen
  \bibfield  {author} {\bibinfo {author} {\bibfnamefont {A}~\bibnamefont
  {Vourdas}},\ }\bibfield  {title} {\enquote {\bibinfo {title} {Analytic
  representations in quantum mechanics},}\ }\href {\doibase
  10.1088/0305-4470/39/7/r01} {\bibfield  {journal} {\bibinfo  {journal}
  {Journal of Physics A: Mathematical and General}\ }\textbf {\bibinfo {volume}
  {39}},\ \bibinfo {pages} {R65–R141} (\bibinfo {year} {2006})}\BibitemShut
  {NoStop}%
\bibitem [{\citenamefont {Holstein}\ and\ \citenamefont
  {Primakoff}(1940)}]{HolsteinPrimakoff}%
  \BibitemOpen
  \bibfield  {author} {\bibinfo {author} {\bibfnamefont {T.}~\bibnamefont
  {Holstein}}\ and\ \bibinfo {author} {\bibfnamefont {H.}~\bibnamefont
  {Primakoff}},\ }\bibfield  {title} {\enquote {\bibinfo {title} {Field
  dependence of the intrinsic domain magnetization of a ferromagnet},}\ }\href
  {\doibase 10.1103/PhysRev.58.1098} {\bibfield  {journal} {\bibinfo  {journal}
  {Phys. Rev.}\ }\textbf {\bibinfo {volume} {58}},\ \bibinfo {pages}
  {1098--1113} (\bibinfo {year} {1940})}\BibitemShut {NoStop}%
\bibitem [{\citenamefont {Honerkamp}\ and\ \citenamefont
  {Hofstetter}(2004)}]{PhysRevLett.92.170403}%
  \BibitemOpen
  \bibfield  {author} {\bibinfo {author} {\bibfnamefont {Carsten}\ \bibnamefont
  {Honerkamp}}\ and\ \bibinfo {author} {\bibfnamefont {Walter}\ \bibnamefont
  {Hofstetter}},\ }\bibfield  {title} {\enquote {\bibinfo {title} {Ultracold
  fermions and the $\mathrm{SU}(n)$ hubbard model},}\ }\href {\doibase
  10.1103/PhysRevLett.92.170403} {\bibfield  {journal} {\bibinfo  {journal}
  {Phys. Rev. Lett.}\ }\textbf {\bibinfo {volume} {92}},\ \bibinfo {pages}
  {170403} (\bibinfo {year} {2004})}\BibitemShut {NoStop}%
\bibitem [{\citenamefont {Zhang}\ \emph {et~al.}(2019)\citenamefont {Zhang},
  \citenamefont {Vidmar},\ and\ \citenamefont {Rigol}}]{PhysRevA.99.063605}%
  \BibitemOpen
  \bibfield  {author} {\bibinfo {author} {\bibfnamefont {Yicheng}\ \bibnamefont
  {Zhang}}, \bibinfo {author} {\bibfnamefont {Lev}\ \bibnamefont {Vidmar}}, \
  and\ \bibinfo {author} {\bibfnamefont {Marcos}\ \bibnamefont {Rigol}},\
  }\bibfield  {title} {\enquote {\bibinfo {title} {Quantum dynamics of
  impenetrable $\mathrm{SU}(n)$ fermions in one-dimensional lattices},}\ }\href
  {\doibase 10.1103/PhysRevA.99.063605} {\bibfield  {journal} {\bibinfo
  {journal} {Phys. Rev. A}\ }\textbf {\bibinfo {volume} {99}},\ \bibinfo
  {pages} {063605} (\bibinfo {year} {2019})}\BibitemShut {NoStop}%
\bibitem [{\citenamefont {Cazalilla}\ \emph {et~al.}(2009)\citenamefont
  {Cazalilla}, \citenamefont {Ho},\ and\ \citenamefont
  {Ueda}}]{Cazalilla_2009}%
  \BibitemOpen
  \bibfield  {author} {\bibinfo {author} {\bibfnamefont {M~A}\ \bibnamefont
  {Cazalilla}}, \bibinfo {author} {\bibfnamefont {A~F}\ \bibnamefont {Ho}}, \
  and\ \bibinfo {author} {\bibfnamefont {M}~\bibnamefont {Ueda}},\ }\bibfield
  {title} {\enquote {\bibinfo {title} {Ultracold gases of ytterbium:
  ferromagnetism and mott states in an {SU}(6) fermi system},}\ }\href
  {\doibase 10.1088/1367-2630/11/10/103033} {\bibfield  {journal} {\bibinfo
  {journal} {New Journal of Physics}\ }\textbf {\bibinfo {volume} {11}},\
  \bibinfo {pages} {103033} (\bibinfo {year} {2009})}\BibitemShut {NoStop}%
\bibitem [{\citenamefont {Laird}\ \emph {et~al.}(2017)\citenamefont {Laird},
  \citenamefont {Shi}, \citenamefont {Parish},\ and\ \citenamefont
  {Levinsen}}]{PhysRevA.96.032701}%
  \BibitemOpen
  \bibfield  {author} {\bibinfo {author} {\bibfnamefont {E.~K.}\ \bibnamefont
  {Laird}}, \bibinfo {author} {\bibfnamefont {Z.-Y.}\ \bibnamefont {Shi}},
  \bibinfo {author} {\bibfnamefont {M.~M.}\ \bibnamefont {Parish}}, \ and\
  \bibinfo {author} {\bibfnamefont {J.}~\bibnamefont {Levinsen}},\ }\bibfield
  {title} {\enquote {\bibinfo {title} {Su($n$) fermions in a one-dimensional
  harmonic trap},}\ }\href {\doibase 10.1103/PhysRevA.96.032701} {\bibfield
  {journal} {\bibinfo  {journal} {Phys. Rev. A}\ }\textbf {\bibinfo {volume}
  {96}},\ \bibinfo {pages} {032701} (\bibinfo {year} {2017})}\BibitemShut
  {NoStop}%
\bibitem [{\citenamefont {Klimov}(2002)}]{Klimov_JMP2002}%
  \BibitemOpen
  \bibfield  {author} {\bibinfo {author} {\bibfnamefont {A.~B.}\ \bibnamefont
  {Klimov}},\ }\bibfield  {title} {\enquote {\bibinfo {title} {Exact evolution
  equations for su(2) quasidistribution functions},}\ }\href {\doibase
  10.1063/1.1463711} {\bibfield  {journal} {\bibinfo  {journal} {Journal of
  Mathematical Physics}\ }\textbf {\bibinfo {volume} {43}},\ \bibinfo {pages}
  {2202--2213} (\bibinfo {year} {2002})}\BibitemShut {NoStop}%
\end{thebibliography}%
\end{document}